\documentclass[twocolumn,aps,pra,superscriptaddress,floatfix]{revtex4-2}

\usepackage{amsmath}
\usepackage{amssymb}
\usepackage{amsfonts}
\usepackage{array}
\usepackage[english]{babel}
\usepackage{bm}
\usepackage{color}
\usepackage{float}
\usepackage{graphicx}
\usepackage{hyperref}
\hypersetup{colorlinks=true, urlcolor=blue, citecolor=blue, linkcolor=red}

\newcommand{\be}{\begin{equation}}
\newcommand{\ee}{\end{equation}}

\definecolor{darkgreen}{rgb}{0, 0.7, 0}

\newcommand\scalemath[2]{\scalebox{#1}{\mbox{\ensuremath{\displaystyle #2}}}}

\newcommand{\PT}{$\mathcal{PT}$}

\begin{document}
	
	 \title{Quantum dynamics of non-Hermitian many-body Landau-Zener systems}

\author{Rajesh K. Malla}
\email{rmalla@bnl.gov}
\affiliation{Condensed Matter Physics and Materials Science Division, Brookhaven National Laboratory, Upton, New York 11973, USA}
\affiliation{Center for Nonlinear Studies, Los Alamos National Laboratory, Los Alamos, New Mexico 87545, USA}
\affiliation{Theoretical Division, Los Alamos National Laboratory, Los Alamos, New Mexico 87545, USA}
\author{Julia Cen}
\email{julia.cen@outlook.com}
\affiliation{Center for Nonlinear Studies, Los Alamos National Laboratory, Los Alamos, New Mexico 87545, USA}
\affiliation{Theoretical Division, Los Alamos National Laboratory, Los Alamos, New Mexico 87545, USA}
\author{Wilton J. M. Kort-Kamp}
\affiliation{Theoretical Division, Los Alamos National Laboratory, Los Alamos, New Mexico 87545, USA}
\author{Avadh Saxena}
\affiliation{Center for Nonlinear Studies, Los Alamos National Laboratory, Los Alamos, New Mexico 87545, USA}
\affiliation{Theoretical Division, Los Alamos National Laboratory, Los Alamos, New Mexico 87545, USA}

	\date{\today}
	
\begin{abstract}
 We develop a framework to solve a large class of linearly driven non-Hermitian quantum systems. Such a class of models in the Hermitian scenario is commonly known as multi-state Landau-Zener models. The non-hermiticity is due to the anti-Hermitian couplings between the diabatic levels. We find that there exists a new conservation law, unique to this class of models, that describes the simultaneous growth of the unnormalized wavefunctions. 
 These models have practical applications in Bose-Einstein condensates, and they can describe the dynamics of multi-species bosonic systems. The conservation law relates to a  pair-production mechanism that explains the dissociation of diatomic molecules into atoms. We provide a general framework for both solvable and semiclassically solvable non-Hermitian Landau-Zener models. Our findings will open new avenues for a number of diverse emergent phenomena in explicitly time-dependent non-Hermitian quantum systems. 

\end{abstract}
	
\maketitle
\section{Introduction}
Non-hermiticity has long been used to understand the dynamics of open quantum systems that cannot be explained by standard Hermitian physics \cite{moiseyev_2011, ashida_2020}. They have gained a lot of attention over the years for their unusual phenomena, such as exceptional points \cite{Heiss_2012, Doppler2016, Jose2016, Peng_2016, Chen_2017_EP, Hodaei_2017, Wang_2021_EP}, non-Hermitian skin effects \cite{Alvarez2018, Kunst2018, Yao_2018, Xiong2018}, and the non-Bloch bulk-boundary correspondence \cite{Yao2018_chern, Yao_2018, Deng_2019, Yokomizo_2019, Kawabata_2020, Yang_2020}. A special class of non-Hermitian systems is also known to exhibit purely real-valued energy eigenspectrum provided they satisfy an antilinear symmetry; a seminal example is joint parity-time ($\mathcal{PT}$)-symmetry \cite{bender_1998, bender_1999}. A consistent quantum mechanical framework may also be constructed for these systems \cite{Mostafazadeh2002,Figueira_2006,FRING20172318}. $\mathcal{PT}$-symmetry has been observed across various areas in physical science, including in optics \cite{El-Ganainy:07, Klaiman_2008, Guo_2009, Ruter_2010_optics}, electronics \cite{Schindler_2011_active, Wang_2020_observation}, mechanics \cite{bender_2013_mechanical}, and acoustics \cite{Zhu_2014_acoustics}. This class of non-Hermiticity has been demonstrated to have a transformational impact in nonreciprocal photonic devices \cite{Ramezani2010, Peng2014} and implications for improving decoherence, entanglement entropy, and Fisher information \cite{Gardas2016, FringFrith2019, cen2022} in quantum information. Despite all these important works in non-Hermitian physics, the key mechanisms driving the dynamics of various explicitly time-dependent non-Hermitian systems outside of the Schr{\"o}dinger picture are poorly understood, and their practical applications remain to be unraveled.

In general, the Schr{\"o}dinger equation with a non-Hermitian Hamiltonian does not enforce energy conservation and, therefore, is not applicable in closed quantum systems. However, in some scenarios, the typically untrackable physics of many-body Hermitian systems can be fully extracted from a simpler single-particle non-Hermitian Hamiltonian.  This is possible due to a one-to-one mapping of the  Schrodinger equation to the Heisenberg equation of motion for bosonic operators via the Bogoliubov-de Gennes (BdG) transformation \cite{BDG} applied to quadratic Hamiltonians.
These many-body Hamiltonians can be made time-dependent by varying the chemical potential of the bosonic modes. The resulting one-particle non-Hermitian time-dependent Hamiltonian predicts the dissociation dynamics of diatomic bosonic molecules (with bosonic atoms) in a mean-field approximation \cite{yurovsky_quantum_2002,kayali}. Interestingly, a many-body generalization of such a non-Hermitian model can be used to investigate the dissociation of a mixture of molecules, given by quadratic many-body Hamiltonians, where the reactions are triggered due to the crossing of chemical potentials \cite{malla2022}. In photonic platforms, such dynamic models can be used to tune anti-Hermitian couplings in photonic waveguides to generate a coherent amplification of light \cite{imag-coupling}.

 Despite growing interest in many-body non-Hermitian dynamics, finding exactly solvable models that enable a comprehensive understanding of the physical mechanisms governing the temporal evolution in these systems remains a challenge to overcome. Even in Hermitian physics, where quantum integrability has been used as an effective tool to produce a variety of solvable time-dependent models \cite{sinitsyn_integrable_2018} with anticipated applications \cite{babujian-22,yuzbashyan-LZ,impulse-LZ,gammam,quest-LZ},  there is still no general recipe for identifying solvable models with a combinatorially large phase space. Remarkably, the analog of quantum integrability in non-Hermitian time-dependent Hamiltonians is still nonexistent. If and how quantum integrability leads to additional solvable non-Hermitian models remains unclear to date.

In this paper, we present a method for constructing exactly solvable, time-dependent many-body non-Hermitian Hamiltonians from known multistate Landau-Zener (MLZ) models \cite{Ryd1,Ryd2,MallaPRL,
altland-LZ,itin,qed1,qed2,qed3,qed4,qed5,qed6,Mallacont} via anti-Hermitian couplings. Recently, other non-Hermitian variants of the Landau-Zener (LZ) model have been explored \cite{Oleh1,Longstaff_2019_LZ, Melanathuru_2022, Kam_2023_LZ}. We revisit the three key phenomena relevant to MLZ models, now explicitly including non-Hermiticity: the independent crossing approximation (ICA), quantum integrability, and the conservation of probability. We observe that there exists a new conservation law that puts a constraint on unnormalized amplitudes. The solutions of the NMLZ models are intertwined with their corresponding MLZ models and can be exactly predicted from the solution of the Hermitian solvable models.

\section{Transition probabilities in non-Hermitian multistate Landau-Zener model}
The non-Hermitian multistate Landau-Zener (NMLZ) Hamiltonians have the form
\be
 {\cal H}(t)= {\cal B} t+{\cal A},
\label{class}
\ee
where ${\cal B}$ and ${\cal A}$ are constant $N  \times N$ matrices and ${\cal B}$ is diagonal. 
The matrix $A$ can be further divided into 
${\cal A}={\cal E}+{\cal G}$,
where ${\cal E}$ is diagonal and describes the static part of the diabatic eigenvalues of ${\cal H}(t)$, and the level couplings are included in the matrix ${\cal G}$. Non-Hermiticity is introduced into ${\cal H}(t)$ via the anti-Hermitian condition, ${\cal G}^{\dagger}=-{\cal G}$.

Our goal is to find solutions for the class of Hamiltonian models described by Eq. (\ref{class}),  {\it i.e.,} to calculate the transition probabilities at  $t\rightarrow +\infty$, when the system evolves from an initial condition at $t\rightarrow -\infty$. Let the wave function $|{\psi}(t)\rangle$, with $N$ amplitudes ${\phi}_{i}(t)$, be the solution of the time-dependent Schr\"{o}dinger equation associated with the Hamiltonian in Eq. (\ref{class}). The spreading of wave function from one diabatic level to $N$ diabatic levels involves many level crossings between diabatic levels. The solution at $t\rightarrow \infty$ requires finding a matrix $S$, which satisfies 
$
|{\psi}(t\rightarrow \infty)\rangle=S |{\psi}(t\rightarrow -\infty)\rangle,
$
where $S\equiv U(T,-T)_{T\rightarrow \infty}$ is a non-unitary matrix of dimension $N$ and $U$ is the time-evolution operator. For a completely solvable model, all the matrix elements in the matrix can be evaluated analytically up to a phase factor.  

The concept of probability is inherent in Hermitian quantum mechanics where the conservation of norm $\langle \psi(t)|\psi(t)\rangle=1$ ensures that the probability of finding a particle in each level is given by $|\phi_n(t)|^2$, which is smaller than or equal to one. In non-Hermitian systems, the norm $\langle \psi(t)|\psi(t)\rangle\neq 1$, since $|\phi_n(t)|^2$ grows in time due to non-unitary evolution. Therefore, we redefine transition probabilities for non-Hermitian wave functions. 
Let us denote the true transition probability from the \emph{n}-th state to the \emph{m}-th state by $P_{mn}$, where  $\sum_{m} P_{mn}=1$. The transition probabilities $P_{mn}$ relate to the unnormalized probabilities,  ${\tilde P}_{mn}=|\phi_{m}(\infty)|^2/|\phi_{n}(-\infty)|^2$ with $\phi_{n}(-\infty)=1$, and are given by
\begin{equation}
    P_{mn}=\frac{{\tilde P}_{mn}}{\sum_{k} {\tilde P}_{kn}}.
    \label{Tp}
\end{equation}
In this paper, we only consider the initial conditions when one of the amplitudes is one, and all others are zero, $\phi_m(-\infty) = \delta_{mn}$, in which case ${\tilde P}_{mn}$ coincides with $|S_{mn}|^2$.

\begin{figure}
\includegraphics[width=0.5\textwidth]{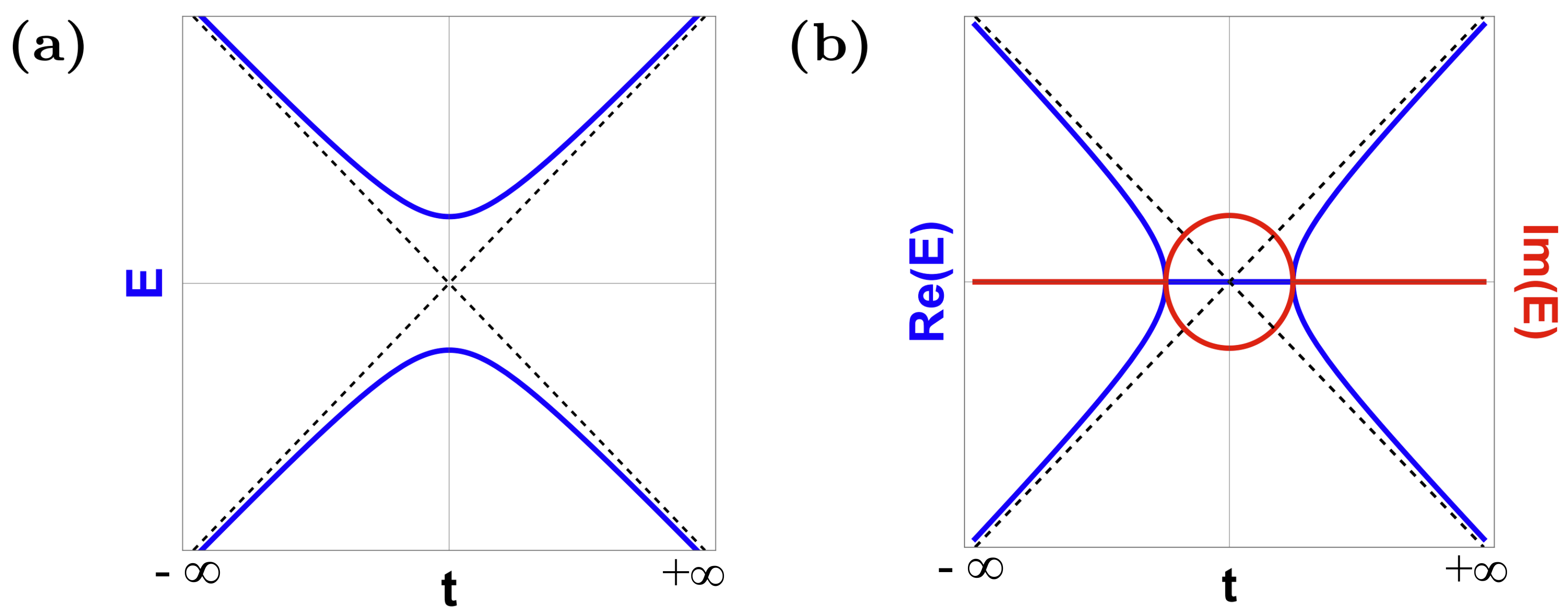}
\caption{Schematic diagrams of the time-dependent eigenvalues of (a) a Hermitian LZ model and (b) a non-Hermitian LZ model Hamiltonian matrix as a function of time. The dashed lines correspond to zero coupling between the two levels. The blue (red) color corresponds to the real (imaginary) part of the eigenvalues.}
\label{fig:1}
\end{figure}
\section{Two-level system}
Let us review the results for the simple case of the $N=2$ NMLZ model, also called the non-Hermitian Landau-Zener (NLZ) model. The Hamiltonian corresponds to the equation of motion 
$i\frac{d}{dt}|{\cal \psi}(t)\rangle={\cal H}_2(t)|{\cal \psi}(t)\rangle$ has the form 
\begin{equation}
{\cal H}_2(t)=\begin{pmatrix}
    -v t && g \\ -g && vt
\end{pmatrix},
    \label{eq:NLZham}
\end{equation}
and represents a two-level system with an anti-Hermitian level coupling, where parameters $v$ and $g$ are real and $\sigma_{z,\pm}$ are the usual Pauli matrices. The instantaneous eigenvalues of ${\cal H}_2(t)$ are  $E_{2,\pm}=\pm\sqrt{(vt)^2-|g|^2}$, which are imaginary in the time interval $|t|<|g|/v$, shown in Fig. \ref{fig:1}b.  The unnormalized probability to remain in the same diabatic level is given by  ${\tilde P}_{11}=e^{2\pi |g|^2/v}$ and the unnormalized probability of transition is  ${\tilde P}_{21}=e^{2\pi |g|^2/v}-1$, see Appendix A. This leads to the relation
\begin{equation}
{\tilde P}_{11}-{\tilde P}_{21}=1.
    \label{cons:NH}
\end{equation}
 The true probabilities are then given by $P_{11}={\tilde P}_{11}/({\tilde P}_{11}+{\tilde P}_{21})$ and  $P_{21}={\tilde P}_{21}/({\tilde P}_{11}+{\tilde P}_{21})$. Similar results have been recently shown in the simplest time-dependent non-Hermitian parity-time (\PT) symmetric variants of the Landau-Zener (LZ) model \cite{Longstaff_2019_LZ}.

 Equation (\ref{cons:NH}) describes the conservation of unnormalized transition probabilities, {\it i.e.,} the rate of growth of the two amplitudes is equal, and the rate is determined by the area under the curve of the imaginary part of the two eigenvalues. The conservation law (\ref{cons:NH}) is of non-Hermitian origin, and has a one-to-one mapping with the pair-production mechanism present in the process of dynamical dissociation, due to Feshbach resonance, of diatomic molecular condensates, in which two atoms are coherently produced \cite{yurovsky_quantum_2002,kayali}, see Appendix B. Note, this mapping only holds in the mean-field approximation when the molecular field operator is replaced by a complex number \cite{yurovsky_quantum_2002}. The Hamiltonian ${\cal H}_2(t)$ can also describe the spontaneous production of bosonic particles when the corresponding field parameters become time-dependent and the evolution passes through
a resonance.  When the number of atomic modes in the system becomes high ($N>2$) to accommodate many degrees of freedom, i.e., spin angular momentum, rotational, and vibrational
modes, then the dynamics can be mapped to the class of Hamiltonians in Eq. (\ref{class}) \cite{malla2022}.

\section{N-level system}
The NMLZ models consisting of $N$ levels can have a maximum of $N(N-1)/2$ level couplings. A complete solution requires finding all $N^2$ transition probabilities. Such a solution, for a generic NMLZ model of class (\ref{class}), does not exist. However, a complete solution may be possible when additional constraints are present within the system. Below, we outline three key phenomena that are relevant to MLZ models, now including non-Hermiticity: the independent crossing approximation (ICA), quantum integrability, and the conservation of probability. 
 
 \subsection{Independent crossing approximation (ICA)} 
 According to ICA, in the Hermitian MLZ model, when a transition probability $P_{mn}$ involves only a single path, $P_{mn}$ can be expressed by a simple application of the two-state LZ formula at every intersection of diabatic energies \cite{Dem, ICA}. For example, 
 certain elements of the $S$ matrix that can be found by Brundobler and Elser (BE) formula for any Hermitian MLZ model (${\cal G}^{\dagger}={\cal G}$), and ${\cal B}_{nm}=b_n \delta_{mn}$, and is given by $S_{nn}=\exp\left(-\pi \sum_{m\neq n}|{\cal G}_{nm}|^2/(|b_{n}-b_{m}|) \right)$, where $|b_n|$ is maximum,  i.e., a diabatic level with maximum slope. 
By following the methodology introduced in Ref. \cite{nogo-LZ}, we find a new BE formula for the scattering matrix element $S_{nn}$ that holds for the class of non-Hermitian Hamiltonians of Eq. (\ref{class}), namely
\begin{equation}
S_{nn}=\exp\left(+\pi \sum_{m\neq n}\frac{|{\cal G}_{nm}|^2}{|b_{n}-b_{m}| }\right).
    \label{modifiedBE}
\end{equation}
 The sign in the exponent of $S_{nn}$ is negative in the MLZ model, while in the NMLZ model, it is positive, see Appendix C. In the Hermitian model, the transition probability ${\tilde P}_{nn}=\prod_{n\neq m} \tilde {p}_{nm}$, where ${\tilde p}_{nm}=\exp\left(-2\pi |{\cal G}_{nm}|^2/|b_{n}-b_{m}| \right)$, while in the non-Hermitian model the unnormalized transition probability ${\tilde P}_{nn}=\prod_{n\neq m} \tilde {p}_{nm}$, where ${\tilde p}_{nm}=\exp\left(+2\pi |{\cal G}_{nm}|^2/|b_{n}-b_{m}| \right)$.
 However, finding partial elements ${\tilde P}_{mn}$ is not enough to know about the true transition probabilities. Unlike Hermitian models,  we need to evaluate all the elements ${\tilde P}_{mn}$ to find the true transition probabilities. 

  \begin{figure}[t!]
    \centering \includegraphics[width=0.45\textwidth]{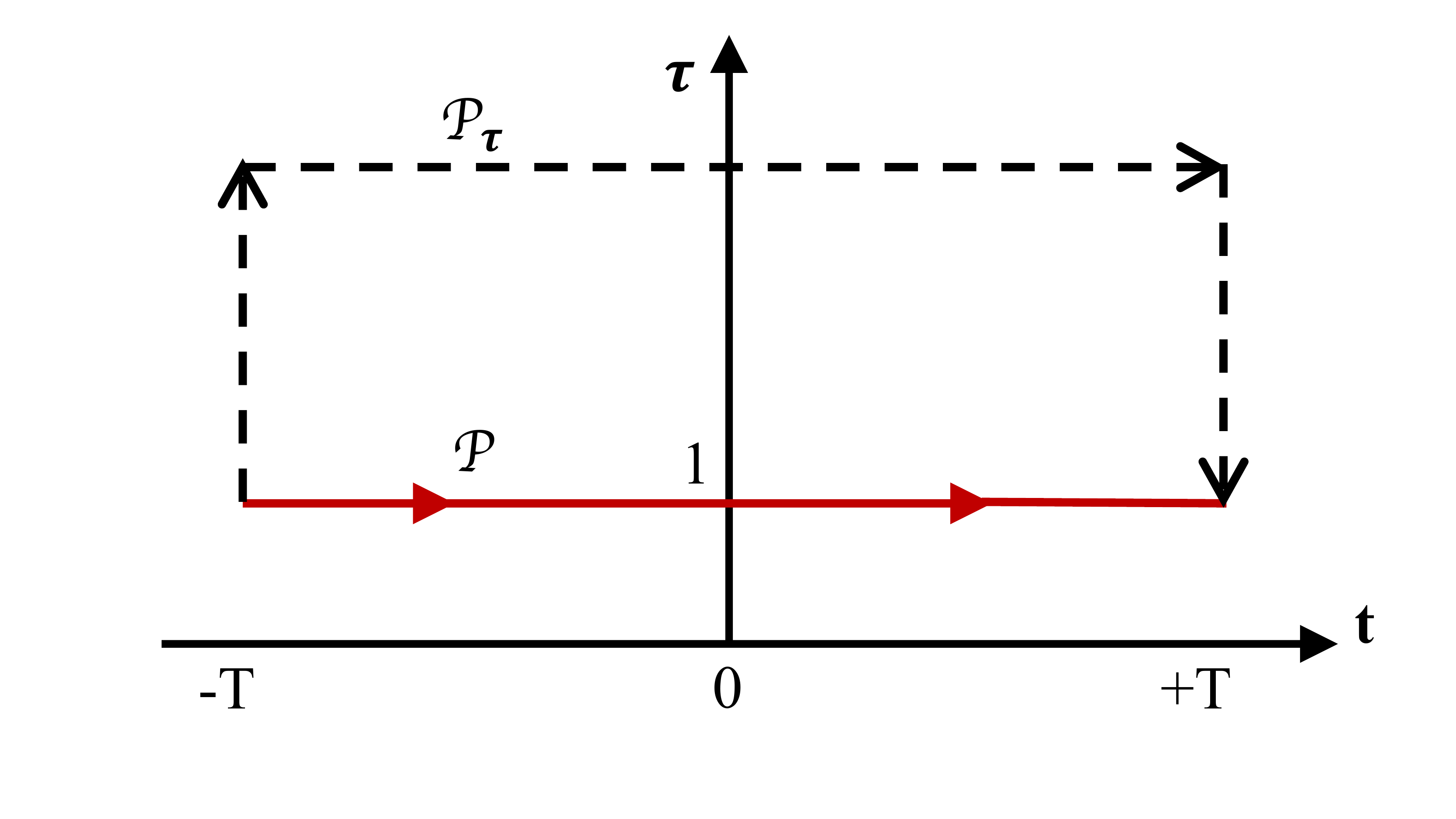}
    \caption{The true time-evolution path ${\cal P}$ (red) with $\tau=1$ and $t\in (-\infty,+\infty)$ can be deformed into the path ${\cal P}_{\tau}$, such that the horizontal part of ${\cal P}_{\tau}$ has $\tau = {\rm constant} \ne 1$ (dashed black arrows). }
    \label{fig:2}%
\end{figure}

\subsection{Integrability of the NMLZ model} 
The theory of integrability of explicitly time-dependent quantum systems has played a key role in finding solutions of MLZ models \cite{sinitsyn_integrable_2018}, predicting dynamical phase transitions in molecular and atomic conversion processes \cite{DPA, MallaPRL}. The integrability condition can be understood as follows. Let us consider a time-dependent Hamiltonian $H(t)$ and the corresponding time evolution operator 
$U=\hat{\cal T}_{\cal P}\exp\left( -i \int_{t_0}^{t_f} H(t)\, dt\right)$
where $\hat{\cal T}_{\cal P}$ is the time ordering operator. Integrability of a Hamiltonian $H(t)$ in time-dependent systems is defined as the possibility of finding a nontrivial Hamiltonian $H'(\tau)$ such that 
\begin{eqnarray}
 \label{cond1}
 \frac{\partial H}{\partial \tau} - \frac{\partial H'}{\partial t}-i[H, H']&=&0.
\end{eqnarray} 
Time evolution corresponding to $H'$ is along $\tau$. Both $H$ and $H'$ are dependent on $t$ and $\tau$. When $H$ and $H'$ satisfy condition (\ref{cond1}), one can deform the integration path of the evolution without changing the amplitudes of the wave functions, and the evolution operator \cite{sinitsyn_integrable_2018}
 \be
 U=\hat{\cal T}_{\cal P}\exp \left( -i \int_{\cal P} H(t,\tau)\, dt+H'(t,\tau) \, d\tau \right),
 \label{path1}
 \ee
where $\hat{\cal T}_{\cal P}$ is the path ordering operator along  ${\cal P}$ in the two-time space $(t,\tau)$, is path independent. One can then transform the physical evolution of the original problem from $t=-\infty$ to $t=\infty$ at a fixed $\tau$, to an evolution along any path in the $(t,\tau)$ plane,  and still achieve the same result for the transition probabilities. 
One such path is to start at $t=-\infty$ and $\tau=1$, evolve along $\tau$, then at fixed $\tau$, evolve along $t$ to $t=\infty$ and finally come back to $\tau=1$ at $t=-\infty$.   The unnormalized transition probability will only depend on the horizontal path; the vertical paths only alter the phase of the wave function, see Fig. \ref{fig:2}.

 The condition in Eq. (\ref{cond1}) further splits into two conditions if the matrices $H$ and $H'$ are purely real,  $\frac{\partial H}{\partial \tau} - \frac{\partial H'}{\partial t}=0$ and $[H, H']=0$.  The condition in Eq. (\ref{cond1}) is not restricted only to Hermitian models, and can be utilized to search for exactly solvable non-Hermitian models. We will demonstrate that 
if an MLZ model $H$ satisfies the condition in Eq.  (\ref{cond1}) with another Hamiltonian $H'$, then the corresponding class of NMLZ models in  Eq.(\ref{class}) constructed from the MLZ model $H$ is also integrable. Note that the difference between the Hermitian and non-Hermitian models is that in the non-Hermitian models, the off-diagonal elements satisfy the anti-Hermiticity condition. So by making the off-diagonal terms anti-Hermitian, one can construct NMLZ models from known MLZ models $H$ and $H'$.
\subsection{Conservation of unnormalized probabilities}
In MLZ models, the amplitudes $\phi_{i}$ satisfy the conservation of probabilities, $\sum_i |\phi_{i}|^2=1$. In the NMLZ model, the true probabilities are calculated only after all the unnormalized probabilities are known. The sum of unnormalized probabilities is not constant and can grow with the system size. We find that there exists a conservation law that describes a pair-production mechanism or a simultaneous growth of wave functions. The conservation law is not universal and depends on the parameters of the NMLZ Hamiltonian as well as the initial conditions. Hence, we will demonstrate the conservation law via different representative examples. This conservation law has a mapping to the unpairing dynamics of the bosonic molecule into atoms, where the conserved quantity is the number of atoms at the start of the reaction \cite{yurovsky_quantum_2002}.

\section{Fully solvable models}
Let ${\cal H}_{sol}$ be an exactly solvable MLZ Hamiltonian. All the  
transition probabilities can be obtained analytically and typically can be expressed as a function of ${\bar p},{\bar q}$, $P_{mn}=f({\bar p},{\bar q})$, where ${\bar p}$ and ${\bar q}$ are two vectors containing parameters corresponding to individual LZ transitions along the path from level $n$ at $t\rightarrow -\infty$ to level $m$ at $t\rightarrow \infty$, and the parameters are $p_{ij}=\exp(-2\pi |{\cal G}_{ij}|^2/|b_i-b_j|)$ and $q_{ij}=1-\exp(-2\pi |{\cal G}_{ij}|^2/|b_i-b_j|)$, respectively. The functional form of $f$ depends on specific problems and the levels considered. 

Now, let us construct an NMLZ model, ${\cal H}_{sol}^{N}$ from ${\cal H}_{sol}$. We claim that ${\cal H}_{sol}^{N}$ is also exactly solvable and the unnormalized transition probabilities ${\tilde P}_{mn}=f({\bar {\tilde p}},{\bar {\tilde q}})$ have the same functional form as their Hermitian counterparts and the vectors ${\bar {\tilde p}}$, and ${\bar {\tilde q}}$, where ${\tilde p}_{ij}=\exp(2\pi |{\cal G}_{ij}|^2/|b_i-b_j|)$ and ${\tilde q}_{ij}=\exp(2\pi |{\cal G}_{ij}|^2/|b_i-b_j|)-1$. It is straightforward to see that the modified BE formula Eq. (\ref{modifiedBE}) keeps the same functional form as the standard BE formula. This claim is indeed general and applies to all the elements of matrix ${\tilde P}$. We illustrate this via two examples below.
\subsection{Example of a N=4 NMLZ model}
Let us consider an exactly solvable $N=4$ NMLZ model constructed from an exactly solvable MLZ Hamiltonian that describes electron tunneling in a double quantum dot \cite{sinitsyn_exact_2015}, which has the form 

\begin{equation}
{\hat {\cal H}}_4
=\begin{pmatrix}
    b_1 t+E_1 & 0 & g^* & -\gamma^*\\0 & -b_1 t+E_1 & \gamma^* & g^*\\
    -g & -\gamma & b_2 t+E_2 & 0 \\\gamma & -g & 0 & -b_2t+E_2
    \end{pmatrix}.
    \label{modelH}
\end{equation}
This example is carried out in the context of molecular dissociation in Ref. \onlinecite{malla2022} and the solution for transition probabilities has the form 
\begin{equation}
{\tilde P}=
\begin{pmatrix}
    \tilde{p}_g \tilde{p}_{\gamma} & 0 & \tilde{p}_{\gamma}\tilde{q}_g& \tilde{q}_{\gamma}\\
    0 & \tilde{p}_g \tilde{p}_{\gamma} & \tilde{q}_{\gamma} & \tilde{p}_{\gamma}\tilde{q}_g \\ \tilde{p}_{\gamma}\tilde{q}_g & \tilde{q}_{\gamma} & \tilde{p}_g \tilde{p}_{\gamma} & 0 \\
    \tilde{q}_{\gamma} & \tilde{p}_{\gamma}\tilde{q}_g& 0 & \tilde{p}_g \tilde{p}_{\gamma}
\end{pmatrix},
    \label{unnormp}
\end{equation}
where $ \tilde{p}_g=e^{\frac{\pi |g|^2}{2|\beta_2|}}, \,\tilde{p}_{\gamma}=e^{\frac{\pi |\gamma|^2}{2|\beta_1|}}$, $2\beta_1=b_1+b_2$, and $2\beta_2=b_1-b_2$. Here, $\tilde{p}_g$ and $\tilde{p}_{\gamma}$ correspond to the NLZ transitions at diabatic level crossings.

The matrix elements ${\tilde P}_{mn}$ in Eq. (\ref{unnormp}) have the same functional form as $P_{mn}$ of the corresponding Hermitian model \cite{sinitsyn_exact_2015}. 
Moreover, we find that there exists a conservation law  
\begin{equation}
    [{\tilde P}_{11}+{\tilde P}_{21}]-[{\tilde P}_{31}+{\tilde P}_{41}]=1,
    \label{cons:1}
\end{equation}
if the initial state $\phi_{1}(-\infty)=1$. This conservation law connects the first column of the matrix ${\tilde P}$; a different initial condition will produce a different conservation law.

\subsection{Example of a N=6 NMLZ model}
\begin{figure}
\includegraphics[width=0.5\textwidth]{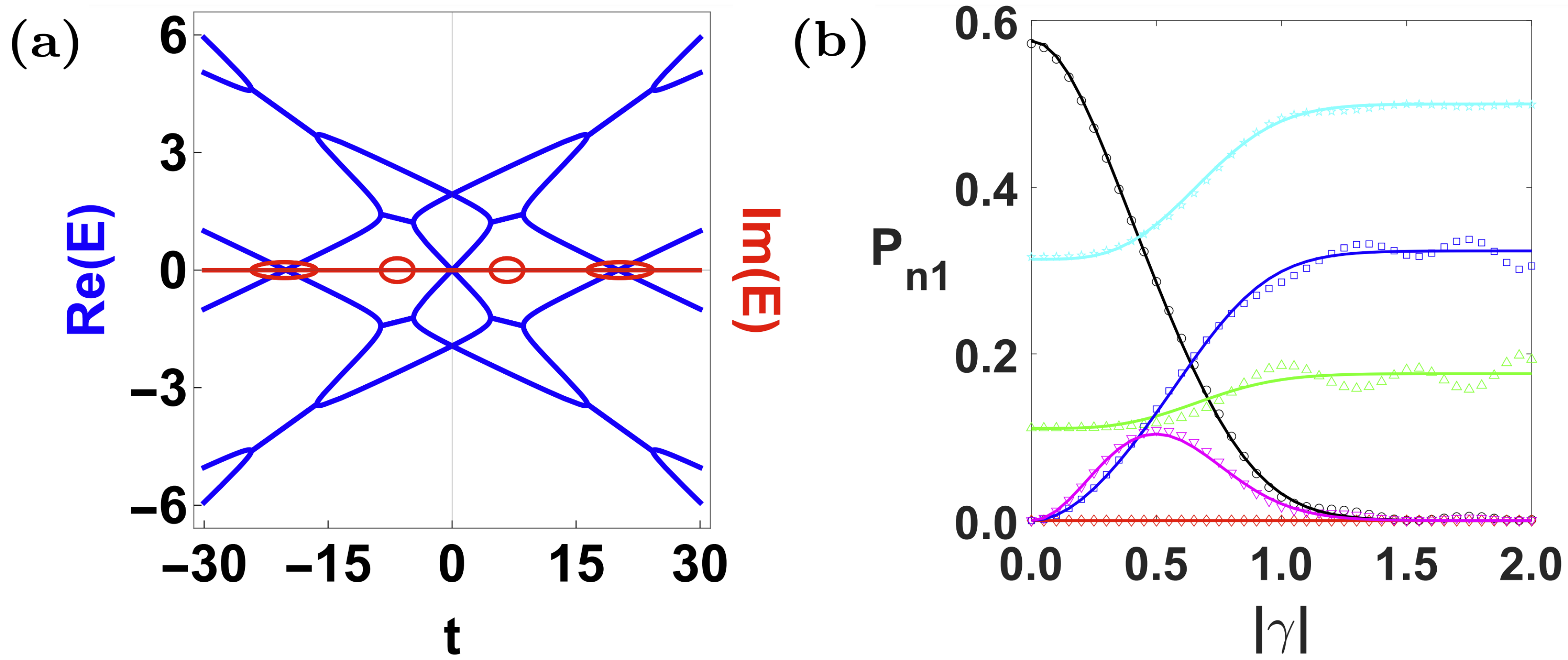}
\caption{Dynamics for the $N=6$ NMLZ model (\ref{modelH6}). (a) The (blue) real and (red) imaginary parts of the eigenvalues of the matrix (\ref{modelH6}) are shown for $E=2$, $b_{1}=0.1$, $b_{2}=0.2$, with couplings $g=0.2$ and $\gamma =0.3$. Parameters $E$ and $g$ have the dimension of $1/t$, while $b$ has the dimension of $1/t^2$. (b) The transition probabilities from the first state to state $n=1$ (black-circle), $n=2$ (green-up triangle), $n=3$ (red-diamond), $n=4$ (blue-square), $n=5$ (magenta-down triangle), $n=6$ (cyan-star), where $E=2.2$, $b_{1}=0.3$, $b_{2}=1.6$, with couplings $g=0.3$ and $\gamma =0.3$.}
\label{fig:3}
\end{figure}

Here, we consider another example of an exactly solvable model of class (\ref{class}) for $N=6$. The Hamiltonian matrix has the form
 \begin{align}
	\hspace{-1mm} H_6(t)
 	\hspace{-1mm}=\hspace{-1mm}\scalemath{0.9}{\begin{pmatrix}
		b_1 t-E & 0 & 0 & 0 & -\gamma & g\\0 & b_1 t+E & 0 & 0 & \gamma & g\\
 		0 & 0 & -b_1 t-E & 0 & g & \gamma \\0 & 0 & 0 & -b_1t+E & g & -\gamma\\ \gamma & -\gamma & -g & -g & -b_2t & 0\\
 		-g & -g & -\gamma & \gamma & 0 & b_2t 
 	\end{pmatrix}}.
 	\label{modelH6}
 \end{align}
 This is an extension of the Hermitian model given in \cite{Sinitsyn_2015_four}, which has been shown to be solvable and all the transition probabilities can be obtained analytically. Here, we show that we can find the transition probabilities for our model (\ref{modelH6}) with the same protocol used for the $N=4$ case.  The eigenvalues of $H_6(t)$ are shown as a function of time in Fig. \ref{fig:3}a. There are $4$ anti-linear-broken regimes where the eigenvalues are complex. However, there are only two coupling parameters $g$ and $\gamma$. All the individual NLZ transitions can be characterized by two terms ${\tilde p}_1$ and ${\tilde p}_2$, and they are given by 
 \begin{eqnarray}
 	{\tilde p}_1 &= e^ \frac{2 \pi \lvert g\rvert^2}{\lvert b_1-b_2 \rvert},~
 	{\tilde p}_2 &= e^\frac{2 \pi \lvert \gamma \rvert^2 }{ \lvert b_1+b_2 \rvert},
 \end{eqnarray}
and ${\tilde q}_1 = {\tilde p}_1-1,~ {\tilde q}_2 = {\tilde p}_2-1$. The unnormalized transition probabilities of the model (\ref{modelH6}), for $b_2>b_1$ are then given by
 \begin{align}
 	 {\tilde P}=\begin{pmatrix}
		{\tilde p}_1{\tilde p}_2 & {\tilde q}_2^2 & 0 & {\tilde p}_2{\tilde q}_1{\tilde q}_2 & {\tilde p}_1{\tilde p}_2{\tilde q}_2 & {\tilde p}_2{\tilde q}_1 \\
		\left( {\tilde p}_2{\tilde q}_1\right)^2 & {\tilde p}_1{\tilde p}_2 & {\tilde p}_2{\tilde q}_2{\tilde q}_1 & 0 & {\tilde q}_2 & {\tilde p}_2^2{\tilde p}_1{\tilde q}_1\\
 		0 & {\tilde p}_2{\tilde q}_2{\tilde q}_1 & {\tilde p}_1{\tilde p}_2 & \left({\tilde p}_2{\tilde q}_1\right)^2 & {\tilde p}_2^2{\tilde p}_1{\tilde q}_1 & {\tilde q}_2 \\
 		{\tilde p}_2{\tilde q}_2{\tilde q}_1 & 0 & {\tilde q}_2^2 & {\tilde p}_1{\tilde p}_2 & {\tilde q}_1{\tilde p}_2 & {\tilde p}_1{\tilde p}_2{\tilde q}_2 \\
 		{\tilde q}_2 & {\tilde p}_1{\tilde p}_2{\tilde q}_2 & {\tilde p}_2{\tilde q}_1 & {\tilde p}_2^2{\tilde p}_1{\tilde q}_1 & \left({\tilde p}_1{\tilde p}_2\right)^2 & 0 \\
 		{\tilde p}_2^2{\tilde p}_1{\tilde q}_1 & {\tilde p}_2{\tilde q}_1 & {\tilde q}_2{\tilde p}_2{\tilde p}_1 & {\tilde q}_2 & 0 & \left({\tilde p}_1{\tilde p}_2\right)^2
 	\end{pmatrix}.
  \label{unnormp6}
 \end{align}
 This matrix has been obtained from \cite{sinitsyn_exact_2015}, where $p_{i}$ and $q_{i}$ are replaced by ${\tilde p}_{i}$ and ${\tilde q}_{i}$. Now, we must obtain the true transition probabilities. However, we notice that the sum of elements in each of the columns in ${\tilde P}$ is not the same. Therefore, we must define 
 a normalization for each column and denote it by ${\cal N}_{i}$, where $i$ is the column index of matrix ${\tilde P}$. The analytical result agrees well with the numerical evolution as shown in Fig. \ref{fig:3}b. 

\section{Not-fully solvable model}
\begin{figure}
\hspace{-0.2in}
\includegraphics[width=0.49\textwidth]{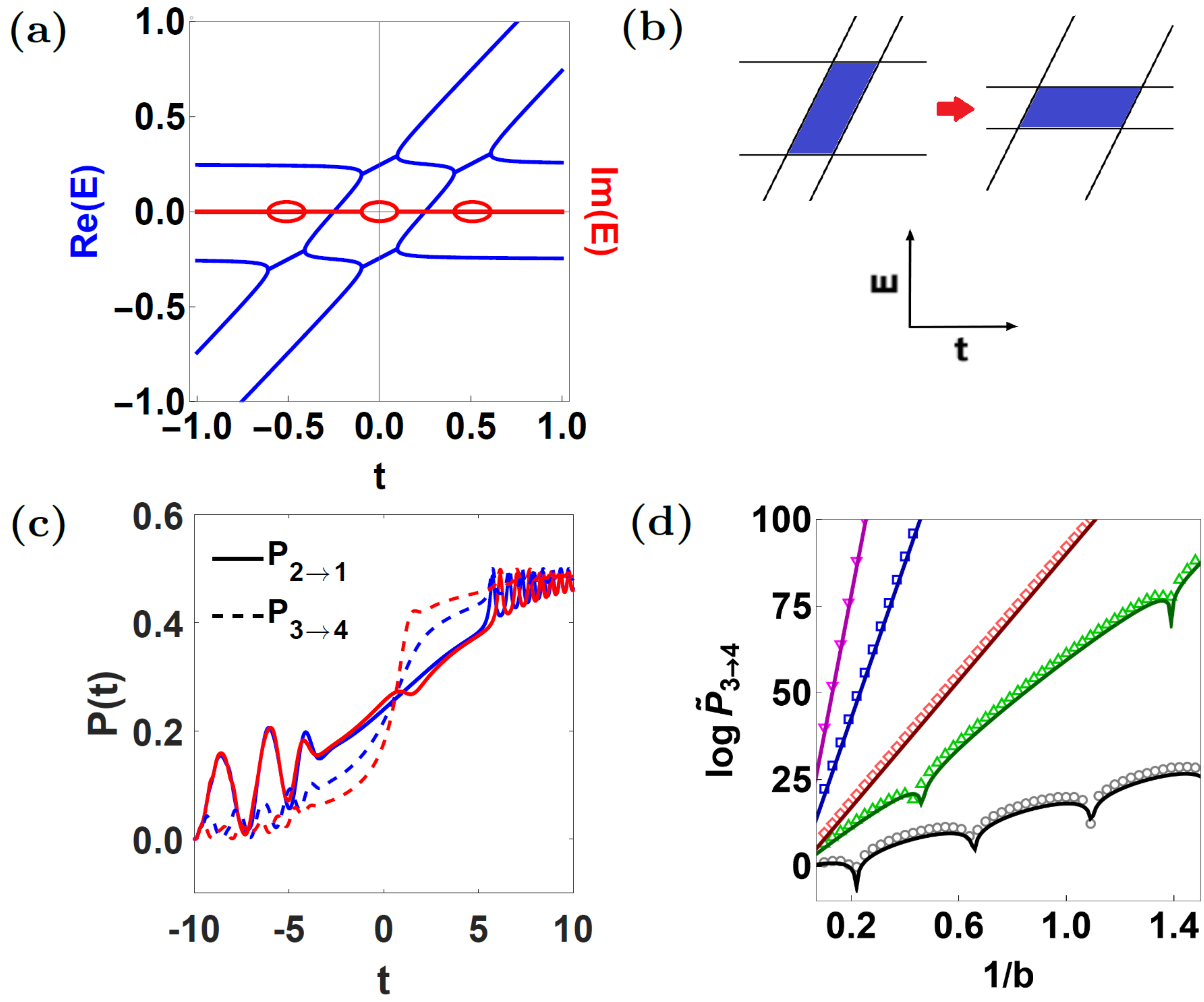}
    \caption{(a) Eigenvalues of N=4 NMLZ model in Eq. (\ref{HS4}) as a function of time with $E_{1}=E_{2}=0.25$, $g=0.05$ and $b=1$. The blue (red) color corresponds to the real (imaginary) part of the eigenvalues. (b) Varying the distances between parallel levels does not change the state-to-state transition probabilities if the blue area enclosed by the diabatic levels is conserved. (c) Numerical normalized transition probabilities for $b=2$, $g=2$, $E_{1}=1$ and $E_{2}=2$, $E_{2}=3$ for the blue and red lines, respectively. (d) The unnormalized transition probabilities $\Tilde{P}_{3\rightarrow 4}$ are shown for the numerical simulations (plot markers) and compared to the analytical approximations (solid lines), where  $E_{1}=E_{2}=2$, with couplings $g^2=1$ (black-circle), $g^2=2\sqrt{2}$ (green-up triangle), $g^2=4$ (red-diamond), $g^2=9$ (blue-square) and $g^2=16$ (magenta-down triangle).}
\label{fig:4}
\end{figure}

 Next, we solve an example NMLZ model constructed from an MLZ model whose exact solution is not known.  Such systems typically host a finite amount of path interferences and some of the transition probabilities depend on the path interference. The Hamiltonian we consider is of the form

\begin{equation}
H_4^S(t)
=\begin{pmatrix}
    E_1 & 0 & g & g\\0 & -E_1 & g & g\\
    -g & -g & b t+E_2 & 0 \\-g & -g & 0 & b t-E_2
    \end{pmatrix},
    \label{HS4}
\end{equation}
which includes two parallel levels crossing another two parallel levels, and the strength of level couplings is considered to be symmetric, as shown in Fig. \ref{fig:4}a. The corresponding MLZ model describes the physics of shuttling electrons in a double quantum-dot system \cite{petta1,petta2, Malla2017nonint}. We identify the two unnormalized transition probabilities, ${\tilde P}_{12}={\tilde P}_{43}$, that are nontrivial (see Fig. \ref{fig:4}c). The remaining unnormalized transition probabilities can be computed using the methods prescribed in the solution of the exactly solvable model. The six unnormalized transition probabilities are ${\tilde P}_{21}={\tilde P}_{34}=0$ \cite{Volkov_2005}, and 
${\tilde P}_{nn}=e^{4\pi|g|^2/b}.$

To find a solution for ${\tilde P}_{43}$ we rely on the symmetry of quantum integrability. 
Our model Eq. (\ref{HS4}) belongs to a class of matrices
\begin{equation}
 H(t,\tau)={\cal B}(\tau) t+E(\tau) {\cal I} +{\cal A}(\tau).
    \label{mlz-tau1}
\end{equation}
Matrices $B(\tau)$ and $A(\tau)$ are obtained from the original $B$ and $A$ by setting
\be
B(\tau) \equiv B\tau, \quad E_1(\tau)\equiv \tau E_1,\quad G(\tau) \equiv G \sqrt{\tau},
\label{tau-intr1}
\ee
and keeping  $E_2$ intact. The corresponding $H'(t,\tau)$ then has the form \cite{malla2021}
\be
H'(t,\tau) = \frac{\partial_{\tau}B(\tau)t^2}{2} +\partial_{\tau} A(\tau)t -\frac{1}{2(b_2-b_1)\tau^2} A^2(\tau).
\label{commute}
\ee
 Another trivial symmetry that appears in all MLZ models is the scaling of time in the Schr{\"o}dinger equation. For example, if we rescale the time $t\rightarrow t/\sqrt{\tau}$, the transition probabilities in the MLZ model remain unchanged. This rescaling corresponds to the changes in parameters in the model (\ref{HS4}),
\begin{equation}
 b_{1,2}\rightarrow b_{1,2}/\tau, \quad E_{1,2} \rightarrow  E_{1,2}/\sqrt{\tau}, \quad G \rightarrow  G/\sqrt{\tau}.
 \label{resc-2}
 \end{equation}
 Under simple inspection, one can see that these transformations should also hold for NMLZ models of class (\ref{class}).
 The simultaneous transformation of (\ref{tau-intr1}) and (\ref{resc-2}) leads to an effective transformation
\begin{equation}
E_1\rightarrow E_1 \sqrt{\tau}, \quad E_2 \rightarrow E_2 /\sqrt{\tau}. 
\label{invar2}
\end{equation}
The physical interpretation of the transformation is shown in Fig. \ref{fig:4}b, where a fixed $\tau$ transforms the level separation between two parallel levels, however, the area under the curve (shown in blue shade) is constant. So, the unnormalized transition probabilities will be invariant as long as the area under the curve is constant. Note, at $\tau=1$ we recover our model Eq. (\ref{HS4}) from $H_4^{S}(t,\tau)$. The advantage of quantum integrability lies in the tunability of parameter $\tau$, {\it i.e.,} the transition probabilities are independent of $\tau$, as presented in Fig. \ref{fig:4}b.

Next, let us express the Schr{\"o}dinger equation for the matrix (\ref{HS4}) as
\begin{eqnarray} i\dot{a_1}&=&E_1a_1+g(a_3+a_4),\label{a1}\\
    i\dot{a_2}&=&-E_1a_2+g(a_3+a_4),\label{a2}\\
    i\dot{a_3}&=&bt+E_2a_3-g(a_1+a_2),\label{a3}\\
    i\dot{a_4}&=&bt-E_2a_4-g(a_1+a_2),\label{a4}
\end{eqnarray}
where $a_1,a_2,a_3,a_4$ are the amplitudes of the levels $1,2,3,4$, respectively. To compute the nontrivial unnormalized transition probability ${\tilde P}_{43}$, we assume that level 3 is initially occupied. Then, we introduce symmetric and anti-symmetric modes $a_{\pm}=(a_1\pm a_2)/{\sqrt{2}}$, and  the equations (\ref{a1}), (\ref{a2}), (\ref{a3}), and (\ref{a4}) transform to 
\begin{eqnarray}
    i\dot{a_+}&=&E_1a_-+\sqrt{2}g(a_3+a_4),\label{aa1}\\
    i\dot{a_-}&=&-E_1a_+,\label{aa2}\\
    i\dot{a_3}&=&(bt+E_2)a_3-\sqrt{2}g(a_+),\label{aa3}\\
    i\dot{a_4}&=&(bt-E_2)a_4-\sqrt{2}g(a_+).\label{aa4}
\end{eqnarray}
Now, we can take advantage of integrability and transform the parameters according to (\ref{tau-intr1}), and set $\tau\rightarrow \infty$. In this limit, levels 3 and 4 cross the symmetric and anti-symmetric modes at $t_{\pm}=\pm E_2/b$ instantly, where ``$+$" corresponds to the crossing of level 4 and ``$-$" corresponds to the crossing of level 3. 

The transition from 3 to 4 can now be understood as a combination of three processes: transition from level 3 to level ``+" near the vicinity of $t_-$, then the dynamics between level ``+" and level ``$-$", then at $t_+$, the transition from level ``+" to level 4. The transitions from and to the symmetric level are individual NLZ transitions where the slope is given by $b\tau$ and the coupling is given by $|\sqrt{2}g\sqrt{\tau}|$. The transitions at $t_{-}$ and $t_{+}$ are simple NLZ transitions and the expression for unnormalized transition probabilities $\tilde{P}_{3\rightarrow +}$ and $\tilde{P}_{+\rightarrow -}$ reads
\begin{equation}
    \tilde{P}_{3\rightarrow +}=\tilde{P}_{+\rightarrow 4}=e^{\frac{4\pi g^2}{b}}-1,
\end{equation}
with the result independent of $\tau$. Now the unnormalized transition probability $\tilde{P}_{3\rightarrow 4}$ can be expressed as
\begin{equation}
    \tilde{P}_{3\rightarrow 4}=\tilde{P}_{3\rightarrow +}\tilde{P}_{+\rightarrow +}\tilde{P}_{+\rightarrow 4},
\end{equation}
where $\tilde{P}_{+\rightarrow +}$ is the unnormalized probability to remain in ``$+$" level after the dynamics with a 2 $\times$ 2 effective Hamiltonian, that acts in the subspace
of $|\pm \rangle$ during the time interval $t \in \{t_{-},t_{+}\}$, see details in \cite{Chernyak_2021}. Away from $t_{\pm}$, the virtual transitions between ``$+$" and ``$-$" can be obtained perturbatively \cite{Chernyak_2021},  and the effective Hamiltonian is given by
\begin{equation}
H_{\rm eff}^s (t) =\frac{1}{b} \left( 
\begin{array}{cc}
\frac{4g^2t}{t^2-1} & e_1e_2 \\
e_1e_2 & 0
\end{array}
\right).
\label{heff-sym}
\end{equation}
Equation (\ref{heff-sym}) describes the Hermitian dynamics and the unnormalized transition probability ${\tilde P}_{++}=P_{++}$ can be obtained by a semiclassical approach \cite{malla2021}. 

The transition probability $P_{++}$ depends on the parameter $r=E_1E_2/|g|^2$. There are two phases $r<1$ and $r>1$, which represent two different behaviors and are given by
\begin{equation}
P_{++}\large =e^{-(2/b) {\rm Im} \left[ \int_{0}^{t_1} \Delta E(t) \, dt \right]}, \quad r<1,
\label{dykhne1-ppp1}
\end{equation}
and 
\begin{equation}
    P_{++}=\Big \vert e^{-\frac{i}{b}  \int_{0}^{t_1} \Delta E(t) \, dt +i\phi_g}+e^{-\frac{i}{b} \int_{0}^{t_2} \Delta E(t) \, dt } \Big \vert^2,\quad r>1,
\label{dykhne1-ppp2}
\end{equation}
where $\Delta E(t)$ is the difference in the eigenvalues of matrix (\ref{heff-sym}) and $t_{1,2}$ are the solutions of equation $\Delta E(t)=0$ close to the real time axis. For  $r<1$, the branching points $t_{1,2}$ are purely imaginary, and the expression for the transition probability $P_{++}$ is estimated with the standard Dykhne formula \cite{dykhne}.  For $r>1$, both the branching points have real as well as imaginary parts, and the imaginary parts are equal to each other. The transition probability (\ref{dykhne1-ppp2}) is a generalized form of the standard Dykhne formula \cite{malla2021}.

The unnormalized transition probability ${\tilde P}_{43}$ is shown in Fig. \ref{fig:4}d as a function of $1/b$ for five values of $r$, which agrees well with our analytical formula. 
The unnormalized transition probability ${\tilde P}_{13}$ can be obtained if one knows ${\tilde P}_{33}$, ${\tilde P}_{43}$, and ${\tilde P}_{23}$. The unnormalized probabilities satisfy the conservation law
\begin{equation}
{\tilde P}_{33}+{\tilde P}_{43}-{\tilde P}_{23}-{\tilde P}_{13}=1.
\end{equation}
This conservation law is similar to that in Eq. (\ref{cons:1}) and depends on the initial condition. The solution to the model in Eq. (\ref{HS4}) is not trivial like the exactly solvable models. A complete solution here requires a combination of Hermitian and non-Hermitian dynamics to fully incorporate all the relevant features. The Hermitian dynamics for the model (\ref{HS4}) captures the effect of the interference, which is absent in the exactly solvable models.

\section{Conclusion}
In conclusion, we have developed a framework to solve a large class of linearly driven non-Hermitian quantum systems. We distinguish between the solution for an exactly solvable model and a not-fully solvable model.
We found that an exactly solvable model has a one-to-one mapping with the respective Hermitian model and demonstrated it via two example models. For a not fully solvable model the solution becomes highly nontrivial. A path to an analytical solution involves a combination of Hermitian and non-Hermitian dynamics. Concerning applications, the pair-production mechanism in our model naturally emerges in the chemical reaction when diatomic molecules undergo a dissociation process in the presence of an external drive.  Moreover, such NMLZ models could potentially be exploited in photonic systems since they provide robust platforms to implement non-Hermiticity.


\section{Acknowledgments}
We would like to thank Andreas Fring and Nikolai Sinitsyn for useful discussions and comments. We gratefully acknowledge the support of the U.S. Department of Energy, the LANL LDRD program, and the Center for Nonlinear Studies at Los Alamos National Laboratory. Part of this work was done at Brookhaven National Laboratory where the work was supported by the U.S. Department of Energy, Office of Basic Energy Sciences, under Contract No. DE-SC0012704.

\appendix
\section{The Non-Hermitian Landau-Zener Model}
Both Hermitian and Non-Hermitian Landau-Zener (LZ) models are described by $2\times 2$ matrices of the form
\begin{equation}
    {\cal H}_2^{(\pm)}(t)=\begin{pmatrix}
        -vt & g\\
        \pm g^* & vt
    \end{pmatrix},
    \label{2level}
\end{equation}
where ``+" refers to the Hermitian and ``$-$" refers to the non-Hermitian model. The eigenvalues of the non-Hermitian matrix are given in Fig. \ref{fig:1}b and are presented alongside the eigenvalues for the standard Hermitian LZ model.  

 The solution of the Schr{\"o}dinger equation with the matrix (\ref{2level}) has the form of a $2\times 1$ column vector,
$$
|\phi(t)\rangle=\begin{pmatrix}
a(t) \\ b(t)
\end{pmatrix}, 
$$
where $a(t)$ satisfies a second order differential equation
\begin{equation}
\ddot{a}(t)+(v^2t^2\pm |g|^2+iv)a(t)=0,
    \label{second}
\end{equation}
whose solutions are given by the parabolic cylinder functions \cite{parabolic,malla2017}. With this, the solution of the Schr{\"o}dinger equation can be expressed as follows 
\begin{equation}
|\phi(t)\rangle=\phi_1\begin{pmatrix}
D_{\nu}(z) \\ -i\sqrt{\nu}D_{\nu-1}(z)
\end{pmatrix}+\phi_2\begin{pmatrix}
D_{\nu}(-z) \\ -i\sqrt{\nu}D_{\nu-1}(-z)
\end{pmatrix},
    \label{linear-comb}
\end{equation}
where $D_{\nu}(z)$ is the parabolic cylinder function, with $\nu=\mp i|g|^2/2\beta$ and $z=\sqrt{2\beta}e^{i\pi/4}t$. 
The difference between the Hermitian and the non-Hermitian dynamics comes from the phase of $\nu$, which is $-\pi/2$ for the Hermitian case and $\pi/2$ for the non-Hermitian case. We are only interested in the asymptotic solution at large times. Assuming the system starts in the upper state, $|a(t\rightarrow -\infty)|^2=1$, the asymptotic solution of $a(t)$ at large positive times is given by $|a(t\rightarrow \infty)|^2=e^{-\pi|g|^2/\beta}$ for the Hermitian model and $|a(t\rightarrow \infty)|^2=e^{\pi|g|^2/\beta}$ for the non-Hermitian model. Similarly, the solution $b(t)$ at large positive times is given by $|b(t\rightarrow \infty)|^2=1-e^{-\pi|g|^2/\beta}$ for the Hermitian model and $|b(t\rightarrow \infty)|^2=e^{\pi|g|^2/\beta}-1$ for the non-Hermitian model.

\section{Relation of NLZ model to molecular dissociation in mean-field approximation}
The simple dissociation process of a molecular BEC into two-mode atomic condensate is $AB\rightarrow A+B$. The Hamiltonian describing such a chemical process is given by \cite{yurovsky_quantum_2002,kayali}
\be
{\hat H}_{2}=\mu_1(t){\hat a}^{\dagger}{\hat a}+\mu_2(t){\hat b}^{\dagger}{\hat b}+J{\hat \psi}^{\dagger}{\hat a}{\hat b}+J^*{\hat \psi}{\hat a}^{\dagger}{\hat b}^{\dagger},
\label{curve-1}
\ee
where  $\mu_i(t)$ are the time-dependent chemical potentials of atomic modes, and the chemical potential of molecular mode can be rescaled to zero. In the nonadiabatic limit, only a small fraction of molecules get converted into atoms \cite{altland-LZ}. One can then replace the molecular field operator with expectation value $\langle {\hat \psi}\rangle$ and we obtain a system of two interacting atomic modes
\be
{\hat H}_{eff}(t)=\mu_1(t){\hat a}^{\dagger}{\hat a}+\mu_2(t){\hat b}^{\dagger}{\hat b}+g{\hat a}{\hat b}+g^*{\hat a}^{\dagger}{\hat b}^{\dagger}.
\label{curve-2}
\ee
Model (\ref{curve-2}) can be solved in the Heisenberg picture where the operators ${\hat a}(t)$ and ${\hat b}(t)$ satisfy   
\begin{align}
i \frac{d}{dt} \begin{pmatrix}
    {\hat a}(t)\\{\hat b}^{\dagger}(t)\end{pmatrix}=\begin{pmatrix} \mu_1(t) & g^* \\
-g & \mu_2(t)
\end{pmatrix}
 \begin{pmatrix}
    {\hat a}(t)\\{\hat b}^{\dagger}(t)\end{pmatrix}.
    \label{curve-3}
\end{align}
The molecular dissociation occurs near the crossing of two chemical potentials. When the chemical potentials are driven linearly, the matrix in (\ref{curve-3}) resembles the Hamiltonian in Eq. (\ref{eq:NLZham}). 

\section{Derivation of the Modified Brundobler and Elser formula }
\begin{figure}
\centering
  \includegraphics[width=.4\textwidth]{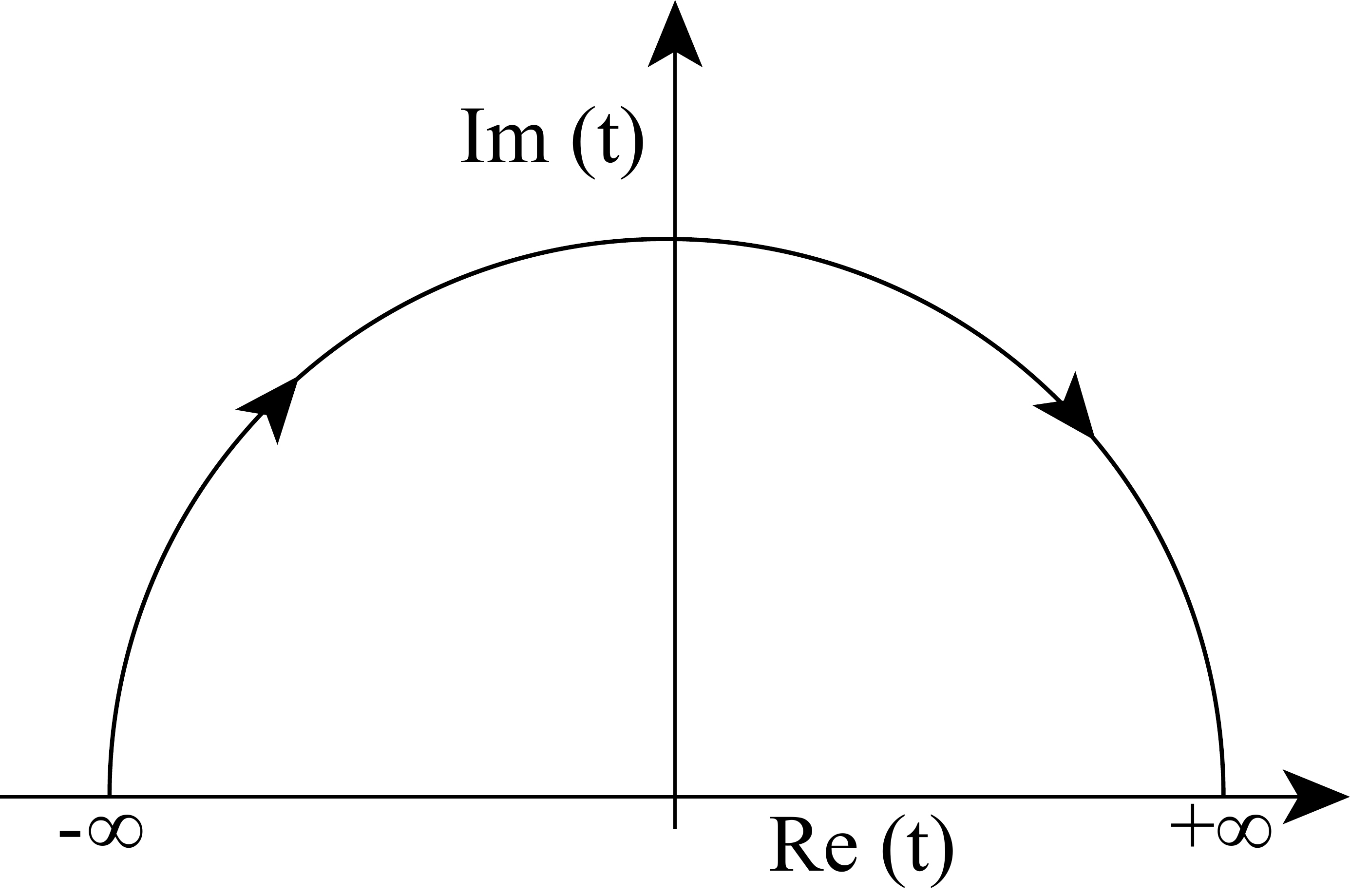}
\caption{Time contour for the evolution from large negative to large positive times
with $t = R \exp(i\phi)$, $R\rightarrow\infty$, $0<\phi<\pi$.}
\label{fig:5}
\end{figure}

The goal of our article is to find the solution of the Schr{\" o}dinger equation corresponding to the class of non-Hermitian matrices (\ref{class}) at asymptotically large times $|t|\rightarrow \infty$. We follow the prescription detailed for Hermitian systems in [\onlinecite{nogo-LZ}], extending the evolution into the complex plane and choosing the evolution path to $|t|\rightarrow \infty$, see Fig. \ref{fig:5}. For small couplings, $|B_{ii}-B_{jj}|t \gg |{\cal G}_{ij}|$, the instantaneous eigenvalues of the matrix remain large for $i \neq j$, and therefore we can use the adiabatic approximation
\begin{equation}
\psi_{i}(t_f) \sim \exp\left(-i\int\limits_{t_i}^{t_f} \epsilon_{i}(t) dt\right) \psi_{i}(t_i),
    \label{adiabatic}
\end{equation}
where the state $\psi_{i}$ has the leading asymptotic form $\psi_{i}\left(t \right) \sim \exp{\left(-i B_{ii}t^{2}/2\right)}$ at $t\rightarrow -\infty$.

Let us consider the state $|0\rangle$ having the highest slope and crossing many states $|i\rangle$. Equation (\ref{adiabatic}) becomes exact at large times. The energy of the state $|0\rangle$ can be expressed up to the first order of $1/|t|$ and is given by~
\begin{equation}
    \epsilon_{0}(t)\sim {\cal E}_{00}-\sum_{i} \frac{|{\cal G}_{i0}|^2}{|B_{00}-B_{ii}|t}.
    \label{energy}
\end{equation}
Substituting (\ref{energy}) into equation (\ref{adiabatic}), we arrive at equation (\ref{modifiedBE}).

\bibliographystyle{apsrev4-2}
\bibliography{RefPRL}

\begin{thebibliography}{78}%
\makeatletter
\providecommand \@ifxundefined [1]{%
 \@ifx{#1\undefined}
}%
\providecommand \@ifnum [1]{%
 \ifnum #1\expandafter \@firstoftwo
 \else \expandafter \@secondoftwo
 \fi
}%
\providecommand \@ifx [1]{%
 \ifx #1\expandafter \@firstoftwo
 \else \expandafter \@secondoftwo
 \fi
}%
\providecommand \natexlab [1]{#1}%
\providecommand \enquote  [1]{``#1''}%
\providecommand \bibnamefont  [1]{#1}%
\providecommand \bibfnamefont [1]{#1}%
\providecommand \citenamefont [1]{#1}%
\providecommand \href@noop [0]{\@secondoftwo}%
\providecommand \href [0]{\begingroup \@sanitize@url \@href}%
\providecommand \@href[1]{\@@startlink{#1}\@@href}%
\providecommand \@@href[1]{\endgroup#1\@@endlink}%
\providecommand \@sanitize@url [0]{\catcode `\\12\catcode `\$12\catcode `\&12\catcode `\#12\catcode `\^12\catcode `\_12\catcode `\%12\relax}%
\providecommand \@@startlink[1]{}%
\providecommand \@@endlink[0]{}%
\providecommand \url  [0]{\begingroup\@sanitize@url \@url }%
\providecommand \@url [1]{\endgroup\@href {#1}{\urlprefix }}%
\providecommand \urlprefix  [0]{URL }%
\providecommand \Eprint [0]{\href }%
\providecommand \doibase [0]{https://doi.org/}%
\providecommand \selectlanguage [0]{\@gobble}%
\providecommand \bibinfo  [0]{\@secondoftwo}%
\providecommand \bibfield  [0]{\@secondoftwo}%
\providecommand \translation [1]{[#1]}%
\providecommand \BibitemOpen [0]{}%
\providecommand \bibitemStop [0]{}%
\providecommand \bibitemNoStop [0]{.\EOS\space}%
\providecommand \EOS [0]{\spacefactor3000\relax}%
\providecommand \BibitemShut  [1]{\csname bibitem#1\endcsname}%
\let\auto@bib@innerbib\@empty
\bibitem [{\citenamefont {Moiseyev}(2011)}]{moiseyev_2011}%
  \BibitemOpen
  \bibfield  {author} {\bibinfo {author} {\bibfnamefont {N.}~\bibnamefont {Moiseyev}},\ }\href {https://doi.org/10.1017/CBO9780511976186} {\emph {\bibinfo {title} {Non-Hermitian Quantum Mechanics}}}\ (\bibinfo  {publisher} {Cambridge University Press},\ \bibinfo {year} {2011})\BibitemShut {NoStop}%
\bibitem [{\citenamefont {Ashida}\ \emph {et~al.}(2020)\citenamefont {Ashida}, \citenamefont {Gong},\ and\ \citenamefont {Ueda}}]{ashida_2020}%
  \BibitemOpen
  \bibfield  {author} {\bibinfo {author} {\bibfnamefont {Y.}~\bibnamefont {Ashida}}, \bibinfo {author} {\bibfnamefont {Z.}~\bibnamefont {Gong}},\ and\ \bibinfo {author} {\bibfnamefont {M.}~\bibnamefont {Ueda}},\ }\href {https://doi.org/10.1080/00018732.2021.1876991} {\bibfield  {journal} {\bibinfo  {journal} {Advances in Physics}\ }\textbf {\bibinfo {volume} {69}},\ \bibinfo {pages} {249} (\bibinfo {year} {2020})}\BibitemShut {NoStop}%
\bibitem [{\citenamefont {Heiss}(2012)}]{Heiss_2012}%
  \BibitemOpen
  \bibfield  {author} {\bibinfo {author} {\bibfnamefont {W.~D.}\ \bibnamefont {Heiss}},\ }\href {https://doi.org/10.1088/1751-8113/45/44/444016} {\bibfield  {journal} {\bibinfo  {journal} {Journal of Physics A: Mathematical and Theoretical}\ }\textbf {\bibinfo {volume} {45}},\ \bibinfo {pages} {444016} (\bibinfo {year} {2012})}\BibitemShut {NoStop}%
\bibitem [{\citenamefont {Doppler}\ \emph {et~al.}(2016)\citenamefont {Doppler}, \citenamefont {Mailybaev}, \citenamefont {B\"{o}hm}, \citenamefont {Kuhl}, \citenamefont {Girschik}, \citenamefont {Libisch}, \citenamefont {Milburn}, \citenamefont {Rabl}, \citenamefont {Moiseyev},\ and\ \citenamefont {Rotter}}]{Doppler2016}%
  \BibitemOpen
  \bibfield  {author} {\bibinfo {author} {\bibfnamefont {J.}~\bibnamefont {Doppler}}, \bibinfo {author} {\bibfnamefont {A.~A.}\ \bibnamefont {Mailybaev}}, \bibinfo {author} {\bibfnamefont {J.}~\bibnamefont {B\"{o}hm}}, \bibinfo {author} {\bibfnamefont {U.}~\bibnamefont {Kuhl}}, \bibinfo {author} {\bibfnamefont {A.}~\bibnamefont {Girschik}}, \bibinfo {author} {\bibfnamefont {F.}~\bibnamefont {Libisch}}, \bibinfo {author} {\bibfnamefont {T.~J.}\ \bibnamefont {Milburn}}, \bibinfo {author} {\bibfnamefont {P.}~\bibnamefont {Rabl}}, \bibinfo {author} {\bibfnamefont {N.}~\bibnamefont {Moiseyev}},\ and\ \bibinfo {author} {\bibfnamefont {S.}~\bibnamefont {Rotter}},\ }\href {https://doi.org/https://doi.org/10.1038/nature18605} {\bibfield  {journal} {\bibinfo  {journal} {Nature}\ }\textbf {\bibinfo {volume} {537}},\ \bibinfo {pages} {76–79} (\bibinfo {year} {2016})}\BibitemShut {NoStop}%
\bibitem [{\citenamefont {San-Jose}\ \emph {et~al.}(2016)\citenamefont {San-Jose}, \citenamefont {Cayao}, \citenamefont {Prada},\ and\ \citenamefont {Aguado}}]{Jose2016}%
  \BibitemOpen
  \bibfield  {author} {\bibinfo {author} {\bibfnamefont {P.}~\bibnamefont {San-Jose}}, \bibinfo {author} {\bibfnamefont {J.}~\bibnamefont {Cayao}}, \bibinfo {author} {\bibfnamefont {E.}~\bibnamefont {Prada}},\ and\ \bibinfo {author} {\bibfnamefont {R.}~\bibnamefont {Aguado}},\ }\href {https://doi.org/https://doi.org/10.1038/srep21427} {\bibfield  {journal} {\bibinfo  {journal} {Scientific Reports}\ }\textbf {\bibinfo {volume} {6}},\ \bibinfo {pages} {21427} (\bibinfo {year} {2016})}\BibitemShut {NoStop}%
\bibitem [{\citenamefont {Peng}\ \emph {et~al.}(2016)\citenamefont {Peng}, \citenamefont {\c{S} K.~\"{O}zdemir}, \citenamefont {Liertzer}, \citenamefont {Chen}, \citenamefont {Kramer}, \citenamefont {lmaz}, \citenamefont {Wiersig}, \citenamefont {Rotter},\ and\ \citenamefont {Yang}}]{Peng_2016}%
  \BibitemOpen
  \bibfield  {author} {\bibinfo {author} {\bibfnamefont {B.}~\bibnamefont {Peng}}, \bibinfo {author} {\bibnamefont {\c{S} K.~\"{O}zdemir}}, \bibinfo {author} {\bibfnamefont {M.}~\bibnamefont {Liertzer}}, \bibinfo {author} {\bibfnamefont {W.}~\bibnamefont {Chen}}, \bibinfo {author} {\bibfnamefont {J.}~\bibnamefont {Kramer}}, \bibinfo {author} {\bibfnamefont {H.~Y.}\ \bibnamefont {lmaz}}, \bibinfo {author} {\bibfnamefont {J.}~\bibnamefont {Wiersig}}, \bibinfo {author} {\bibfnamefont {S.}~\bibnamefont {Rotter}},\ and\ \bibinfo {author} {\bibfnamefont {L.}~\bibnamefont {Yang}},\ }\href {https://doi.org/https://doi.org/10.1073/pnas.1603318113} {\bibfield  {journal} {\bibinfo  {journal} {Proceedings of the National academy of Sciences}\ }\textbf {\bibinfo {volume} {113}},\ \bibinfo {pages} {6845} (\bibinfo {year} {2016})}\BibitemShut {NoStop}%
\bibitem [{\citenamefont {Chen}\ \emph {et~al.}(2017)\citenamefont {Chen}, \citenamefont {\c{S}. K.~\"{O}zdemir}, \citenamefont {Zhao}, \citenamefont {Wiersig},\ and\ \citenamefont {Yang}}]{Chen_2017_EP}%
  \BibitemOpen
  \bibfield  {author} {\bibinfo {author} {\bibfnamefont {W.}~\bibnamefont {Chen}}, \bibinfo {author} {\bibnamefont {\c{S}. K.~\"{O}zdemir}}, \bibinfo {author} {\bibfnamefont {G.}~\bibnamefont {Zhao}}, \bibinfo {author} {\bibfnamefont {J.}~\bibnamefont {Wiersig}},\ and\ \bibinfo {author} {\bibfnamefont {L.}~\bibnamefont {Yang}},\ }\href {https://doi.org/https://doi.org/10.1038/nature23281} {\bibfield  {journal} {\bibinfo  {journal} {Nature}\ }\textbf {\bibinfo {volume} {548}},\ \bibinfo {pages} {192–196} (\bibinfo {year} {2017})}\BibitemShut {NoStop}%
\bibitem [{\citenamefont {Hodaei}\ \emph {et~al.}(2017)\citenamefont {Hodaei}, \citenamefont {Hassan}, \citenamefont {Wittek}, \citenamefont {Garcia-Gracia}, \citenamefont {El-Ganainy}, \citenamefont {Christodoulides},\ and\ \citenamefont {Khajavikhan}}]{Hodaei_2017}%
  \BibitemOpen
  \bibfield  {author} {\bibinfo {author} {\bibfnamefont {H.}~\bibnamefont {Hodaei}}, \bibinfo {author} {\bibfnamefont {A.~U.}\ \bibnamefont {Hassan}}, \bibinfo {author} {\bibfnamefont {S.}~\bibnamefont {Wittek}}, \bibinfo {author} {\bibfnamefont {H.}~\bibnamefont {Garcia-Gracia}}, \bibinfo {author} {\bibfnamefont {R.}~\bibnamefont {El-Ganainy}}, \bibinfo {author} {\bibfnamefont {D.~N.}\ \bibnamefont {Christodoulides}},\ and\ \bibinfo {author} {\bibfnamefont {M.}~\bibnamefont {Khajavikhan}},\ }\href {https://doi.org/https://doi.org/10.1038/nature23280} {\bibfield  {journal} {\bibinfo  {journal} {Nature}\ }\textbf {\bibinfo {volume} {548}},\ \bibinfo {pages} {187–191} (\bibinfo {year} {2017})}\BibitemShut {NoStop}%
\bibitem [{\citenamefont {Wang}\ \emph {et~al.}(2021)\citenamefont {Wang}, \citenamefont {Sweeney}, \citenamefont {Stone},\ and\ \citenamefont {Yang}}]{Wang_2021_EP}%
  \BibitemOpen
  \bibfield  {author} {\bibinfo {author} {\bibfnamefont {C.}~\bibnamefont {Wang}}, \bibinfo {author} {\bibfnamefont {W.~R.}\ \bibnamefont {Sweeney}}, \bibinfo {author} {\bibfnamefont {A.~D.}\ \bibnamefont {Stone}},\ and\ \bibinfo {author} {\bibfnamefont {L.}~\bibnamefont {Yang}},\ }\href {https://doi.org/10.1126/science.abj1028} {\bibfield  {journal} {\bibinfo  {journal} {Science}\ }\textbf {\bibinfo {volume} {373}},\ \bibinfo {pages} {1261} (\bibinfo {year} {2021})}\BibitemShut {NoStop}%
\bibitem [{\citenamefont {Alvarez}\ \emph {et~al.}(2018)\citenamefont {Alvarez}, \citenamefont {Vargas}, \citenamefont {Berdakin},\ and\ \citenamefont {Torres}}]{Alvarez2018}%
  \BibitemOpen
  \bibfield  {author} {\bibinfo {author} {\bibfnamefont {V.~M.~M.}\ \bibnamefont {Alvarez}}, \bibinfo {author} {\bibfnamefont {J.~E.~B.}\ \bibnamefont {Vargas}}, \bibinfo {author} {\bibfnamefont {M.}~\bibnamefont {Berdakin}},\ and\ \bibinfo {author} {\bibfnamefont {L.~E. F.~F.}\ \bibnamefont {Torres}},\ }\href {https://doi.org/https://doi.org/10.1140/epjst/e2018-800091-5} {\bibfield  {journal} {\bibinfo  {journal} {The European Physical Journal Special Topics}\ }\textbf {\bibinfo {volume} {227}},\ \bibinfo {pages} {1295–1308} (\bibinfo {year} {2018})}\BibitemShut {NoStop}%
\bibitem [{\citenamefont {Kunst}\ \emph {et~al.}(2018)\citenamefont {Kunst}, \citenamefont {Edvardsson}, \citenamefont {Budich},\ and\ \citenamefont {Bergholtz}}]{Kunst2018}%
  \BibitemOpen
  \bibfield  {author} {\bibinfo {author} {\bibfnamefont {F.~K.}\ \bibnamefont {Kunst}}, \bibinfo {author} {\bibfnamefont {E.}~\bibnamefont {Edvardsson}}, \bibinfo {author} {\bibfnamefont {J.~C.}\ \bibnamefont {Budich}},\ and\ \bibinfo {author} {\bibfnamefont {E.~J.}\ \bibnamefont {Bergholtz}},\ }\href {https://doi.org/10.1103/PhysRevLett.121.026808} {\bibfield  {journal} {\bibinfo  {journal} {Physical Review Letters}\ }\textbf {\bibinfo {volume} {121}},\ \bibinfo {pages} {026808} (\bibinfo {year} {2018})}\BibitemShut {NoStop}%
\bibitem [{\citenamefont {Yao}\ and\ \citenamefont {Wang}(2018)}]{Yao_2018}%
  \BibitemOpen
  \bibfield  {author} {\bibinfo {author} {\bibfnamefont {S.}~\bibnamefont {Yao}}\ and\ \bibinfo {author} {\bibfnamefont {Z.}~\bibnamefont {Wang}},\ }\href {https://doi.org/10.1103/PhysRevLett.121.086803} {\bibfield  {journal} {\bibinfo  {journal} {Physical Review Letters}\ }\textbf {\bibinfo {volume} {121}},\ \bibinfo {pages} {086803} (\bibinfo {year} {2018})}\BibitemShut {NoStop}%
\bibitem [{\citenamefont {Xiong}(2018)}]{Xiong2018}%
  \BibitemOpen
  \bibfield  {author} {\bibinfo {author} {\bibfnamefont {Y.}~\bibnamefont {Xiong}},\ }\href {https://doi.org/https://doi.org/10.1088/2399-6528/aab64a} {\bibfield  {journal} {\bibinfo  {journal} {Journal of Physics Communications}\ }\textbf {\bibinfo {volume} {2}},\ \bibinfo {pages} {035043} (\bibinfo {year} {2018})}\BibitemShut {NoStop}%
\bibitem [{\citenamefont {Yao}\ \emph {et~al.}(2018)\citenamefont {Yao}, \citenamefont {Song},\ and\ \citenamefont {Wang}}]{Yao2018_chern}%
  \BibitemOpen
  \bibfield  {author} {\bibinfo {author} {\bibfnamefont {S.}~\bibnamefont {Yao}}, \bibinfo {author} {\bibfnamefont {F.}~\bibnamefont {Song}},\ and\ \bibinfo {author} {\bibfnamefont {Z.}~\bibnamefont {Wang}},\ }\href {https://doi.org/10.1103/PhysRevLett.121.136802} {\bibfield  {journal} {\bibinfo  {journal} {Physical Review Letters}\ }\textbf {\bibinfo {volume} {121}},\ \bibinfo {pages} {136802} (\bibinfo {year} {2018})}\BibitemShut {NoStop}%
\bibitem [{\citenamefont {Deng}\ and\ \citenamefont {Yi}(2019)}]{Deng_2019}%
  \BibitemOpen
  \bibfield  {author} {\bibinfo {author} {\bibfnamefont {T.-S.}\ \bibnamefont {Deng}}\ and\ \bibinfo {author} {\bibfnamefont {W.}~\bibnamefont {Yi}},\ }\href {https://doi.org/10.1103/PhysRevB.100.035102} {\bibfield  {journal} {\bibinfo  {journal} {Physical Review B}\ }\textbf {\bibinfo {volume} {100}},\ \bibinfo {pages} {035102} (\bibinfo {year} {2019})}\BibitemShut {NoStop}%
\bibitem [{\citenamefont {Yokomizo}\ and\ \citenamefont {Murakami}(2019)}]{Yokomizo_2019}%
  \BibitemOpen
  \bibfield  {author} {\bibinfo {author} {\bibfnamefont {K.}~\bibnamefont {Yokomizo}}\ and\ \bibinfo {author} {\bibfnamefont {S.}~\bibnamefont {Murakami}},\ }\href {https://doi.org/10.1103/PhysRevLett.123.066404} {\bibfield  {journal} {\bibinfo  {journal} {Physical Review Letters}\ }\textbf {\bibinfo {volume} {123}},\ \bibinfo {pages} {066404} (\bibinfo {year} {2019})}\BibitemShut {NoStop}%
\bibitem [{\citenamefont {Kawabata}\ \emph {et~al.}(2020)\citenamefont {Kawabata}, \citenamefont {Okuma},\ and\ \citenamefont {Sato}}]{Kawabata_2020}%
  \BibitemOpen
  \bibfield  {author} {\bibinfo {author} {\bibfnamefont {K.}~\bibnamefont {Kawabata}}, \bibinfo {author} {\bibfnamefont {N.}~\bibnamefont {Okuma}},\ and\ \bibinfo {author} {\bibfnamefont {M.}~\bibnamefont {Sato}},\ }\href {https://doi.org/10.1103/PhysRevB.101.195147} {\bibfield  {journal} {\bibinfo  {journal} {Physical Review B}\ }\textbf {\bibinfo {volume} {101}},\ \bibinfo {pages} {195147} (\bibinfo {year} {2020})}\BibitemShut {NoStop}%
\bibitem [{\citenamefont {Yang}\ \emph {et~al.}(2020)\citenamefont {Yang}, \citenamefont {Zhang}, \citenamefont {Ffang},\ and\ \citenamefont {Hu}}]{Yang_2020}%
  \BibitemOpen
  \bibfield  {author} {\bibinfo {author} {\bibfnamefont {Z.}~\bibnamefont {Yang}}, \bibinfo {author} {\bibfnamefont {K.}~\bibnamefont {Zhang}}, \bibinfo {author} {\bibfnamefont {C.}~\bibnamefont {Ffang}},\ and\ \bibinfo {author} {\bibfnamefont {J.}~\bibnamefont {Hu}},\ }\href {https://doi.org/10.1103/PhysRevLett.125.226402} {\bibfield  {journal} {\bibinfo  {journal} {Physical Review Letters}\ }\textbf {\bibinfo {volume} {125}},\ \bibinfo {pages} {226402} (\bibinfo {year} {2020})}\BibitemShut {NoStop}%
\bibitem [{\citenamefont {Bender}\ and\ \citenamefont {Boettcher}(1998)}]{bender_1998}%
  \BibitemOpen
  \bibfield  {author} {\bibinfo {author} {\bibfnamefont {C.~M.}\ \bibnamefont {Bender}}\ and\ \bibinfo {author} {\bibfnamefont {S.}~\bibnamefont {Boettcher}},\ }\href {https://doi.org/10.1103/PhysRevLett.80.5243} {\bibfield  {journal} {\bibinfo  {journal} {Physical Review Letters}\ }\textbf {\bibinfo {volume} {80}},\ \bibinfo {pages} {5243} (\bibinfo {year} {1998})}\BibitemShut {NoStop}%
\bibitem [{\citenamefont {Bender}\ \emph {et~al.}(1999)\citenamefont {Bender}, \citenamefont {Boettcher},\ and\ \citenamefont {Meisinger}}]{bender_1999}%
  \BibitemOpen
  \bibfield  {author} {\bibinfo {author} {\bibfnamefont {C.~M.}\ \bibnamefont {Bender}}, \bibinfo {author} {\bibfnamefont {S.}~\bibnamefont {Boettcher}},\ and\ \bibinfo {author} {\bibfnamefont {P.~N.}\ \bibnamefont {Meisinger}},\ }\href {https://doi.org/10.1063/1.532860} {\bibfield  {journal} {\bibinfo  {journal} {Journal of Mathematical Physics}\ }\textbf {\bibinfo {volume} {40}},\ \bibinfo {pages} {2201} (\bibinfo {year} {1999})}\BibitemShut {NoStop}%
\bibitem [{\citenamefont {Mostafazadeh}(2002)}]{Mostafazadeh2002}%
  \BibitemOpen
  \bibfield  {author} {\bibinfo {author} {\bibfnamefont {A.}~\bibnamefont {Mostafazadeh}},\ }\href {https://doi.org/10.1063/1.1461427} {\bibfield  {journal} {\bibinfo  {journal} {Journal of Mathematical Physics}\ }\textbf {\bibinfo {volume} {43}},\ \bibinfo {pages} {2814} (\bibinfo {year} {2002})},\ \Eprint {https://arxiv.org/abs/https://pubs.aip.org/aip/jmp/article-pdf/43/5/2814/8171928/2814\_1\_online.pdf} {https://pubs.aip.org/aip/jmp/article-pdf/43/5/2814/8171928/2814\_1\_online.pdf} \BibitemShut {NoStop}%
\bibitem [{\citenamefont {Faria}\ and\ \citenamefont {Fring}(2006)}]{Figueira_2006}%
  \BibitemOpen
  \bibfield  {author} {\bibinfo {author} {\bibfnamefont {C.~F.~M.}\ \bibnamefont {Faria}}\ and\ \bibinfo {author} {\bibfnamefont {A.}~\bibnamefont {Fring}},\ }\href {https://doi.org/10.1088/0305-4470/39/29/018} {\bibfield  {journal} {\bibinfo  {journal} {Journal of Physics A: Mathematical and General}\ }\textbf {\bibinfo {volume} {39}},\ \bibinfo {pages} {9269} (\bibinfo {year} {2006})}\BibitemShut {NoStop}%
\bibitem [{\citenamefont {Fring}\ and\ \citenamefont {Frith}(2017)}]{FRING20172318}%
  \BibitemOpen
  \bibfield  {author} {\bibinfo {author} {\bibfnamefont {A.}~\bibnamefont {Fring}}\ and\ \bibinfo {author} {\bibfnamefont {T.}~\bibnamefont {Frith}},\ }\href {https://doi.org/https://doi.org/10.1016/j.physleta.2017.05.041} {\bibfield  {journal} {\bibinfo  {journal} {Physics Letters A}\ }\textbf {\bibinfo {volume} {381}},\ \bibinfo {pages} {2318} (\bibinfo {year} {2017})}\BibitemShut {NoStop}%
\bibitem [{\citenamefont {El-Ganainy}\ \emph {et~al.}(2007)\citenamefont {El-Ganainy}, \citenamefont {Makris}, \citenamefont {Christodoulides},\ and\ \citenamefont {Musslimani}}]{El-Ganainy:07}%
  \BibitemOpen
  \bibfield  {author} {\bibinfo {author} {\bibfnamefont {R.}~\bibnamefont {El-Ganainy}}, \bibinfo {author} {\bibfnamefont {K.~G.}\ \bibnamefont {Makris}}, \bibinfo {author} {\bibfnamefont {D.~N.}\ \bibnamefont {Christodoulides}},\ and\ \bibinfo {author} {\bibfnamefont {Z.~H.}\ \bibnamefont {Musslimani}},\ }\href {https://doi.org/10.1364/OL.32.002632} {\bibfield  {journal} {\bibinfo  {journal} {Optics Letters}\ }\textbf {\bibinfo {volume} {32}},\ \bibinfo {pages} {2632} (\bibinfo {year} {2007})}\BibitemShut {NoStop}%
\bibitem [{\citenamefont {Klaiman}\ \emph {et~al.}(2008)\citenamefont {Klaiman}, \citenamefont {G\"{u}nther},\ and\ \citenamefont {Moiseyev}}]{Klaiman_2008}%
  \BibitemOpen
  \bibfield  {author} {\bibinfo {author} {\bibfnamefont {S.}~\bibnamefont {Klaiman}}, \bibinfo {author} {\bibfnamefont {U.}~\bibnamefont {G\"{u}nther}},\ and\ \bibinfo {author} {\bibfnamefont {N.}~\bibnamefont {Moiseyev}},\ }\href {https://doi.org/10.1103/PhysRevLett.101.080402} {\bibfield  {journal} {\bibinfo  {journal} {Physical Review Letters}\ }\textbf {\bibinfo {volume} {101}},\ \bibinfo {pages} {080402} (\bibinfo {year} {2008})}\BibitemShut {NoStop}%
\bibitem [{\citenamefont {Guo}\ \emph {et~al.}(2009)\citenamefont {Guo}, \citenamefont {Salamo}, \citenamefont {Duchesne}, \citenamefont {Morandotti}, \citenamefont {Volatier-Ravat}, \citenamefont {Aimez}, \citenamefont {Siviloglou},\ and\ \citenamefont {Christodoulides}}]{Guo_2009}%
  \BibitemOpen
  \bibfield  {author} {\bibinfo {author} {\bibfnamefont {A.}~\bibnamefont {Guo}}, \bibinfo {author} {\bibfnamefont {G.~J.}\ \bibnamefont {Salamo}}, \bibinfo {author} {\bibfnamefont {D.}~\bibnamefont {Duchesne}}, \bibinfo {author} {\bibfnamefont {R.}~\bibnamefont {Morandotti}}, \bibinfo {author} {\bibfnamefont {M.}~\bibnamefont {Volatier-Ravat}}, \bibinfo {author} {\bibfnamefont {V.}~\bibnamefont {Aimez}}, \bibinfo {author} {\bibfnamefont {G.~A.}\ \bibnamefont {Siviloglou}},\ and\ \bibinfo {author} {\bibfnamefont {D.~N.}\ \bibnamefont {Christodoulides}},\ }\href {https://doi.org/10.1103/PhysRevLett.103.093902} {\bibfield  {journal} {\bibinfo  {journal} {Physical Review Letters}\ }\textbf {\bibinfo {volume} {103}},\ \bibinfo {pages} {093902} (\bibinfo {year} {2009})}\BibitemShut {NoStop}%
\bibitem [{\citenamefont {R\"{u}ter}\ \emph {et~al.}(2010)\citenamefont {R\"{u}ter}, \citenamefont {Makris}, \citenamefont {El-Ganainy}, \citenamefont {Christodoulides}, \citenamefont {Segev},\ and\ \citenamefont {Kip}}]{Ruter_2010_optics}%
  \BibitemOpen
  \bibfield  {author} {\bibinfo {author} {\bibfnamefont {C.~E.}\ \bibnamefont {R\"{u}ter}}, \bibinfo {author} {\bibfnamefont {K.~G.}\ \bibnamefont {Makris}}, \bibinfo {author} {\bibfnamefont {R.}~\bibnamefont {El-Ganainy}}, \bibinfo {author} {\bibfnamefont {D.~N.}\ \bibnamefont {Christodoulides}}, \bibinfo {author} {\bibfnamefont {M.}~\bibnamefont {Segev}},\ and\ \bibinfo {author} {\bibfnamefont {D.}~\bibnamefont {Kip}},\ }\href {https://doi.org/https://doi.org/10.1038/nphys1515} {\bibfield  {journal} {\bibinfo  {journal} {Nature Physics}\ }\textbf {\bibinfo {volume} {6}},\ \bibinfo {pages} {192} (\bibinfo {year} {2010})}\BibitemShut {NoStop}%
\bibitem [{\citenamefont {Schindler}\ \emph {et~al.}(2011)\citenamefont {Schindler}, \citenamefont {Li}, \citenamefont {Zheng}, \citenamefont {Ellis},\ and\ \citenamefont {Kottos}}]{Schindler_2011_active}%
  \BibitemOpen
  \bibfield  {author} {\bibinfo {author} {\bibfnamefont {J.}~\bibnamefont {Schindler}}, \bibinfo {author} {\bibfnamefont {A.}~\bibnamefont {Li}}, \bibinfo {author} {\bibfnamefont {M.~C.}\ \bibnamefont {Zheng}}, \bibinfo {author} {\bibfnamefont {F.~M.}\ \bibnamefont {Ellis}},\ and\ \bibinfo {author} {\bibfnamefont {T.}~\bibnamefont {Kottos}},\ }\href {https://doi.org/10.1103/PhysRevA.84.040101} {\bibfield  {journal} {\bibinfo  {journal} {Physical Review A}\ }\textbf {\bibinfo {volume} {84}},\ \bibinfo {pages} {040101} (\bibinfo {year} {2011})}\BibitemShut {NoStop}%
\bibitem [{\citenamefont {Wang}\ \emph {et~al.}(2020)\citenamefont {Wang}, \citenamefont {Fang}, \citenamefont {Xie}, \citenamefont {Dong}, \citenamefont {Joglekar}, \citenamefont {Wang}, \citenamefont {Li},\ and\ \citenamefont {Luo}}]{Wang_2020_observation}%
  \BibitemOpen
  \bibfield  {author} {\bibinfo {author} {\bibfnamefont {T.}~\bibnamefont {Wang}}, \bibinfo {author} {\bibfnamefont {J.}~\bibnamefont {Fang}}, \bibinfo {author} {\bibfnamefont {Z.}~\bibnamefont {Xie}}, \bibinfo {author} {\bibfnamefont {N.}~\bibnamefont {Dong}}, \bibinfo {author} {\bibfnamefont {Y.~N.}\ \bibnamefont {Joglekar}}, \bibinfo {author} {\bibfnamefont {Z.}~\bibnamefont {Wang}}, \bibinfo {author} {\bibfnamefont {J.}~\bibnamefont {Li}},\ and\ \bibinfo {author} {\bibfnamefont {L.}~\bibnamefont {Luo}},\ }\href {https://doi.org/10.1140/epjd/e2020-10131-7} {\bibfield  {journal} {\bibinfo  {journal} {The European Physical Journal D}\ }\textbf {\bibinfo {volume} {74}} (\bibinfo {year} {2020})}\BibitemShut {NoStop}%
\bibitem [{\citenamefont {Bender}\ \emph {et~al.}(2013)\citenamefont {Bender}, \citenamefont {Berntson}, \citenamefont {Parker},\ and\ \citenamefont {Samuel}}]{bender_2013_mechanical}%
  \BibitemOpen
  \bibfield  {author} {\bibinfo {author} {\bibfnamefont {C.~M.}\ \bibnamefont {Bender}}, \bibinfo {author} {\bibfnamefont {B.~K.}\ \bibnamefont {Berntson}}, \bibinfo {author} {\bibfnamefont {D.}~\bibnamefont {Parker}},\ and\ \bibinfo {author} {\bibfnamefont {E.}~\bibnamefont {Samuel}},\ }\href {https://doi.org/https://doi.org/10.1119/1.4789549} {\bibfield  {journal} {\bibinfo  {journal} {American Journal of Physics}\ }\textbf {\bibinfo {volume} {81}},\ \bibinfo {pages} {173} (\bibinfo {year} {2013})}\BibitemShut {NoStop}%
\bibitem [{\citenamefont {Zhu}\ \emph {et~al.}(2014)\citenamefont {Zhu}, \citenamefont {Ramezani}, \citenamefont {Shi}, \citenamefont {Zhu},\ and\ \citenamefont {Zhang}}]{Zhu_2014_acoustics}%
  \BibitemOpen
  \bibfield  {author} {\bibinfo {author} {\bibfnamefont {X.}~\bibnamefont {Zhu}}, \bibinfo {author} {\bibfnamefont {H.}~\bibnamefont {Ramezani}}, \bibinfo {author} {\bibfnamefont {C.}~\bibnamefont {Shi}}, \bibinfo {author} {\bibfnamefont {J.}~\bibnamefont {Zhu}},\ and\ \bibinfo {author} {\bibfnamefont {X.}~\bibnamefont {Zhang}},\ }\href {https://doi.org/10.1103/PhysRevX.4.031042} {\bibfield  {journal} {\bibinfo  {journal} {Physical Review X}\ }\textbf {\bibinfo {volume} {4}},\ \bibinfo {pages} {031042} (\bibinfo {year} {2014})}\BibitemShut {NoStop}%
\bibitem [{\citenamefont {Ramezani}\ \emph {et~al.}(2010)\citenamefont {Ramezani}, \citenamefont {Kottos}, \citenamefont {El-Ganainy},\ and\ \citenamefont {Christodoulides}}]{Ramezani2010}%
  \BibitemOpen
  \bibfield  {author} {\bibinfo {author} {\bibfnamefont {H.}~\bibnamefont {Ramezani}}, \bibinfo {author} {\bibfnamefont {T.}~\bibnamefont {Kottos}}, \bibinfo {author} {\bibfnamefont {R.}~\bibnamefont {El-Ganainy}},\ and\ \bibinfo {author} {\bibfnamefont {D.~N.}\ \bibnamefont {Christodoulides}},\ }\href {https://doi.org/10.1103/PhysRevA.82.043803} {\bibfield  {journal} {\bibinfo  {journal} {Physical Review A}\ }\textbf {\bibinfo {volume} {82}},\ \bibinfo {pages} {043803} (\bibinfo {year} {2010})}\BibitemShut {NoStop}%
\bibitem [{\citenamefont {Peng}\ \emph {et~al.}(2014)\citenamefont {Peng}, \citenamefont {\"{O}zdemir}, \citenamefont {Lei}, \citenamefont {Monifi}, \citenamefont {Gianfreda}, \citenamefont {Long}, \citenamefont {Fan}, \citenamefont {Nori}, \citenamefont {Bender},\ and\ \citenamefont {Yang}}]{Peng2014}%
  \BibitemOpen
  \bibfield  {author} {\bibinfo {author} {\bibfnamefont {B.}~\bibnamefont {Peng}}, \bibinfo {author} {\bibfnamefont {{\c S}.~K.}\ \bibnamefont {\"{O}zdemir}}, \bibinfo {author} {\bibfnamefont {F.}~\bibnamefont {Lei}}, \bibinfo {author} {\bibfnamefont {F.}~\bibnamefont {Monifi}}, \bibinfo {author} {\bibfnamefont {M.}~\bibnamefont {Gianfreda}}, \bibinfo {author} {\bibfnamefont {G.~L.}\ \bibnamefont {Long}}, \bibinfo {author} {\bibfnamefont {S.}~\bibnamefont {Fan}}, \bibinfo {author} {\bibfnamefont {F.}~\bibnamefont {Nori}}, \bibinfo {author} {\bibfnamefont {C.~M.}\ \bibnamefont {Bender}},\ and\ \bibinfo {author} {\bibfnamefont {L.}~\bibnamefont {Yang}},\ }\href {https://doi.org/10.1038/nphys2927} {\bibfield  {journal} {\bibinfo  {journal} {Nature Physics}\ }\textbf {\bibinfo {volume} {10}},\ \bibinfo {pages} {394} (\bibinfo {year} {2014})}\BibitemShut {NoStop}%
\bibitem [{\citenamefont {Gardas}\ \emph {et~al.}(2016)\citenamefont {Gardas}, \citenamefont {Deffner},\ and\ \citenamefont {Saxena}}]{Gardas2016}%
  \BibitemOpen
  \bibfield  {author} {\bibinfo {author} {\bibfnamefont {B.}~\bibnamefont {Gardas}}, \bibinfo {author} {\bibfnamefont {S.}~\bibnamefont {Deffner}},\ and\ \bibinfo {author} {\bibfnamefont {A.}~\bibnamefont {Saxena}},\ }\href {https://doi.org/10.1103/PhysRevA.94.040101} {\bibfield  {journal} {\bibinfo  {journal} {Physical Review A}\ }\textbf {\bibinfo {volume} {94}},\ \bibinfo {pages} {040101} (\bibinfo {year} {2016})}\BibitemShut {NoStop}%
\bibitem [{\citenamefont {Fring}\ and\ \citenamefont {Frith}(2019)}]{FringFrith2019}%
  \BibitemOpen
  \bibfield  {author} {\bibinfo {author} {\bibfnamefont {A.}~\bibnamefont {Fring}}\ and\ \bibinfo {author} {\bibfnamefont {T.}~\bibnamefont {Frith}},\ }\href {https://doi.org/10.1103/PhysRevA.100.010102} {\bibfield  {journal} {\bibinfo  {journal} {Physical Review A}\ }\textbf {\bibinfo {volume} {100}},\ \bibinfo {pages} {010102} (\bibinfo {year} {2019})}\BibitemShut {NoStop}%
\bibitem [{\citenamefont {Cen}\ and\ \citenamefont {Saxena}(2022)}]{cen2022}%
  \BibitemOpen
  \bibfield  {author} {\bibinfo {author} {\bibfnamefont {J.}~\bibnamefont {Cen}}\ and\ \bibinfo {author} {\bibfnamefont {A.}~\bibnamefont {Saxena}},\ }\href {https://doi.org/10.1103/PhysRevA.105.022404} {\bibfield  {journal} {\bibinfo  {journal} {Physical Review A}\ }\textbf {\bibinfo {volume} {105}},\ \bibinfo {pages} {022404} (\bibinfo {year} {2022})}\BibitemShut {NoStop}%
\bibitem [{\citenamefont {Rossignoli}\ and\ \citenamefont {Kowalski}(2005)}]{BDG}%
  \BibitemOpen
  \bibfield  {author} {\bibinfo {author} {\bibfnamefont {R.}~\bibnamefont {Rossignoli}}\ and\ \bibinfo {author} {\bibfnamefont {A.~M.}\ \bibnamefont {Kowalski}},\ }\href {https://doi.org/10.1103/PhysRevA.72.032101} {\bibfield  {journal} {\bibinfo  {journal} {Physical Review A}\ }\textbf {\bibinfo {volume} {72}},\ \bibinfo {pages} {032101} (\bibinfo {year} {2005})}\BibitemShut {NoStop}%
\bibitem [{\citenamefont {Yurovsky}\ \emph {et~al.}(2002)\citenamefont {Yurovsky}, \citenamefont {Ben-Reuven},\ and\ \citenamefont {Julienne}}]{yurovsky_quantum_2002}%
  \BibitemOpen
  \bibfield  {author} {\bibinfo {author} {\bibfnamefont {V.~A.}\ \bibnamefont {Yurovsky}}, \bibinfo {author} {\bibfnamefont {A.}~\bibnamefont {Ben-Reuven}},\ and\ \bibinfo {author} {\bibfnamefont {P.~S.}\ \bibnamefont {Julienne}},\ }\href {https://doi.org/10.1103/PhysRevA.65.043607} {\bibfield  {journal} {\bibinfo  {journal} {Physical Review A}\ }\textbf {\bibinfo {volume} {65}},\ \bibinfo {pages} {043607} (\bibinfo {year} {2002})}\BibitemShut {NoStop}%
\bibitem [{\citenamefont {Kayali}\ and\ \citenamefont {Sinitsyn}(2003)}]{kayali}%
  \BibitemOpen
  \bibfield  {author} {\bibinfo {author} {\bibfnamefont {M.~A.}\ \bibnamefont {Kayali}}\ and\ \bibinfo {author} {\bibfnamefont {N.~A.}\ \bibnamefont {Sinitsyn}},\ }\href {https://doi.org/10.1103/PhysRevA.67.045603} {\bibfield  {journal} {\bibinfo  {journal} {Physical Review A}\ }\textbf {\bibinfo {volume} {67}},\ \bibinfo {pages} {045603} (\bibinfo {year} {2003})}\BibitemShut {NoStop}%
\bibitem [{\citenamefont {Malla}(2022)}]{malla2022}%
  \BibitemOpen
  \bibfield  {author} {\bibinfo {author} {\bibfnamefont {R.~K.}\ \bibnamefont {Malla}},\ }\href {https://doi.org/10.1103/PhysRevA.106.033318} {\bibfield  {journal} {\bibinfo  {journal} {Physical Review A}\ }\textbf {\bibinfo {volume} {106}},\ \bibinfo {pages} {033318} (\bibinfo {year} {2022})}\BibitemShut {NoStop}%
\bibitem [{\citenamefont {Ke}\ \emph {et~al.}(2018)\citenamefont {Ke}, \citenamefont {Zhao}, \citenamefont {Liu}, \citenamefont {Wu}, \citenamefont {Wang},\ and\ \citenamefont {Lu}}]{imag-coupling}%
  \BibitemOpen
  \bibfield  {author} {\bibinfo {author} {\bibfnamefont {S.}~\bibnamefont {Ke}}, \bibinfo {author} {\bibfnamefont {D.}~\bibnamefont {Zhao}}, \bibinfo {author} {\bibfnamefont {Q.}~\bibnamefont {Liu}}, \bibinfo {author} {\bibfnamefont {S.}~\bibnamefont {Wu}}, \bibinfo {author} {\bibfnamefont {B.}~\bibnamefont {Wang}},\ and\ \bibinfo {author} {\bibfnamefont {P.}~\bibnamefont {Lu}},\ }\href {https://doi.org/10.1109/JLT.2018.2814038} {\bibfield  {journal} {\bibinfo  {journal} {Journal of Lightwave Technology}\ }\textbf {\bibinfo {volume} {36}},\ \bibinfo {pages} {2510} (\bibinfo {year} {2018})}\BibitemShut {NoStop}%
\bibitem [{\citenamefont {Sinitsyn}\ \emph {et~al.}(2018)\citenamefont {Sinitsyn}, \citenamefont {Yuzbashyan}, \citenamefont {Chernyak}, \citenamefont {Patra},\ and\ \citenamefont {Sun}}]{sinitsyn_integrable_2018}%
  \BibitemOpen
  \bibfield  {author} {\bibinfo {author} {\bibfnamefont {N.~A.}\ \bibnamefont {Sinitsyn}}, \bibinfo {author} {\bibfnamefont {E.~A.}\ \bibnamefont {Yuzbashyan}}, \bibinfo {author} {\bibfnamefont {V.~Y.}\ \bibnamefont {Chernyak}}, \bibinfo {author} {\bibfnamefont {A.}~\bibnamefont {Patra}},\ and\ \bibinfo {author} {\bibfnamefont {C.}~\bibnamefont {Sun}},\ }\href {https://doi.org/10.1103/PhysRevLett.120.190402} {\bibfield  {journal} {\bibinfo  {journal} {Physical Review Letters}\ }\textbf {\bibinfo {volume} {120}},\ \bibinfo {pages} {190402} (\bibinfo {year} {2018})}\BibitemShut {NoStop}%
\bibitem [{\citenamefont {Sedrakyan}\ and\ \citenamefont {Babujian}(2022)}]{babujian-22}%
  \BibitemOpen
  \bibfield  {author} {\bibinfo {author} {\bibfnamefont {T.~A.}\ \bibnamefont {Sedrakyan}}\ and\ \bibinfo {author} {\bibfnamefont {H.~M.}\ \bibnamefont {Babujian}},\ }\href {https://doi.org/10.1007/JHEP04(2022)039} {\bibfield  {journal} {\bibinfo  {journal} {Journal of High Energy Physics}\ }\textbf {\bibinfo {volume} {39}} (\bibinfo {year} {2022})}\BibitemShut {NoStop}%
\bibitem [{\citenamefont {Yuzbashyan}(2018)}]{yuzbashyan-LZ}%
  \BibitemOpen
  \bibfield  {author} {\bibinfo {author} {\bibfnamefont {E.~A.}\ \bibnamefont {Yuzbashyan}},\ }\href {https://doi.org/https://doi.org/10.1016/j.aop.2018.01.017} {\bibfield  {journal} {\bibinfo  {journal} {Annals of Physics}\ }\textbf {\bibinfo {volume} {392}},\ \bibinfo {pages} {323} (\bibinfo {year} {2018})}\BibitemShut {NoStop}%
\bibitem [{\citenamefont {Suzuki}\ and\ \citenamefont {Nakazato}(2022)}]{impulse-LZ}%
  \BibitemOpen
  \bibfield  {author} {\bibinfo {author} {\bibfnamefont {T.}~\bibnamefont {Suzuki}}\ and\ \bibinfo {author} {\bibfnamefont {H.}~\bibnamefont {Nakazato}},\ }\href {https://doi.org/10.1103/PhysRevA.105.022211} {\bibfield  {journal} {\bibinfo  {journal} {Physical Review A}\ }\textbf {\bibinfo {volume} {105}},\ \bibinfo {pages} {022211} (\bibinfo {year} {2022})}\BibitemShut {NoStop}%
\bibitem [{\citenamefont {Chernyak}\ \emph {et~al.}(2019)\citenamefont {Chernyak}, \citenamefont {Sinitsyn},\ and\ \citenamefont {Sun}}]{gammam}%
  \BibitemOpen
  \bibfield  {author} {\bibinfo {author} {\bibfnamefont {V.~Y.}\ \bibnamefont {Chernyak}}, \bibinfo {author} {\bibfnamefont {N.~A.}\ \bibnamefont {Sinitsyn}},\ and\ \bibinfo {author} {\bibfnamefont {C.}~\bibnamefont {Sun}},\ }\href {https://doi.org/10.1103/PhysRevB.100.224304} {\bibfield  {journal} {\bibinfo  {journal} {Physical Review B}\ }\textbf {\bibinfo {volume} {100}},\ \bibinfo {pages} {224304} (\bibinfo {year} {2019})}\BibitemShut {NoStop}%
\bibitem [{\citenamefont {Sinitsyn}\ and\ \citenamefont {Chernyak}(2017)}]{quest-LZ}%
  \BibitemOpen
  \bibfield  {author} {\bibinfo {author} {\bibfnamefont {N.~A.}\ \bibnamefont {Sinitsyn}}\ and\ \bibinfo {author} {\bibfnamefont {V.~Y.}\ \bibnamefont {Chernyak}},\ }\href {https://doi.org/10.1088/1751-8121/aa6800} {\bibfield  {journal} {\bibinfo  {journal} {Journal of Physics A: Mathematical and Theoretical}\ }\textbf {\bibinfo {volume} {50}},\ \bibinfo {pages} {255203} (\bibinfo {year} {2017})}\BibitemShut {NoStop}%
\bibitem [{\citenamefont {Harmin}\ and\ \citenamefont {Price}(1994)}]{Ryd1}%
  \BibitemOpen
  \bibfield  {author} {\bibinfo {author} {\bibfnamefont {D.~A.}\ \bibnamefont {Harmin}}\ and\ \bibinfo {author} {\bibfnamefont {P.~N.}\ \bibnamefont {Price}},\ }\href {https://doi.org/10.1103/PhysRevA.49.1933} {\bibfield  {journal} {\bibinfo  {journal} {Physical Review A}\ }\textbf {\bibinfo {volume} {49}},\ \bibinfo {pages} {1933} (\bibinfo {year} {1994})}\BibitemShut {NoStop}%
\bibitem [{\citenamefont {Harmin}(1997)}]{Ryd2}%
  \BibitemOpen
  \bibfield  {author} {\bibinfo {author} {\bibfnamefont {D.~A.}\ \bibnamefont {Harmin}},\ }\href {https://doi.org/10.1103/PhysRevA.56.232} {\bibfield  {journal} {\bibinfo  {journal} {Physical Review A}\ }\textbf {\bibinfo {volume} {56}},\ \bibinfo {pages} {232} (\bibinfo {year} {1997})}\BibitemShut {NoStop}%
\bibitem [{\citenamefont {Malla}\ \emph {et~al.}(2022)\citenamefont {Malla}, \citenamefont {Chernyak}, \citenamefont {Sun},\ and\ \citenamefont {Sinitsyn}}]{MallaPRL}%
  \BibitemOpen
  \bibfield  {author} {\bibinfo {author} {\bibfnamefont {R.~K.}\ \bibnamefont {Malla}}, \bibinfo {author} {\bibfnamefont {V.~Y.}\ \bibnamefont {Chernyak}}, \bibinfo {author} {\bibfnamefont {C.}~\bibnamefont {Sun}},\ and\ \bibinfo {author} {\bibfnamefont {N.~A.}\ \bibnamefont {Sinitsyn}},\ }\href {https://doi.org/10.1103/PhysRevLett.129.033201} {\bibfield  {journal} {\bibinfo  {journal} {Physical Review Letters}\ }\textbf {\bibinfo {volume} {129}},\ \bibinfo {pages} {033201} (\bibinfo {year} {2022})}\BibitemShut {NoStop}%
\bibitem [{\citenamefont {Altland}\ \emph {et~al.}(2009)\citenamefont {Altland}, \citenamefont {Gurarie}, \citenamefont {Kriecherbauer},\ and\ \citenamefont {Polkovnikov}}]{altland-LZ}%
  \BibitemOpen
  \bibfield  {author} {\bibinfo {author} {\bibfnamefont {A.}~\bibnamefont {Altland}}, \bibinfo {author} {\bibfnamefont {V.}~\bibnamefont {Gurarie}}, \bibinfo {author} {\bibfnamefont {T.}~\bibnamefont {Kriecherbauer}},\ and\ \bibinfo {author} {\bibfnamefont {A.}~\bibnamefont {Polkovnikov}},\ }\href {https://doi.org/10.1103/PhysRevA.79.042703} {\bibfield  {journal} {\bibinfo  {journal} {Physical Review A}\ }\textbf {\bibinfo {volume} {79}},\ \bibinfo {pages} {042703} (\bibinfo {year} {2009})}\BibitemShut {NoStop}%
\bibitem [{\citenamefont {Itin}\ and\ \citenamefont {T\"orm\"a}(2009)}]{itin}%
  \BibitemOpen
  \bibfield  {author} {\bibinfo {author} {\bibfnamefont {A.~P.}\ \bibnamefont {Itin}}\ and\ \bibinfo {author} {\bibfnamefont {P.}~\bibnamefont {T\"orm\"a}},\ }\href {https://doi.org/10.1103/PhysRevA.79.055602} {\bibfield  {journal} {\bibinfo  {journal} {Physical Review A}\ }\textbf {\bibinfo {volume} {79}},\ \bibinfo {pages} {055602} (\bibinfo {year} {2009})}\BibitemShut {NoStop}%
\bibitem [{\citenamefont {Werther}\ \emph {et~al.}(2019)\citenamefont {Werther}, \citenamefont {Grossmann}, \citenamefont {Huang},\ and\ \citenamefont {Zhao}}]{qed1}%
  \BibitemOpen
  \bibfield  {author} {\bibinfo {author} {\bibfnamefont {M.}~\bibnamefont {Werther}}, \bibinfo {author} {\bibfnamefont {F.}~\bibnamefont {Grossmann}}, \bibinfo {author} {\bibfnamefont {Z.}~\bibnamefont {Huang}},\ and\ \bibinfo {author} {\bibfnamefont {Y.}~\bibnamefont {Zhao}},\ }\href {https://doi.org/https://doi.org/10.1063/1.5096158} {\bibfield  {journal} {\bibinfo  {journal} {The Journal of Chemical Physics}\ }\textbf {\bibinfo {volume} {150}},\ \bibinfo {pages} {234109} (\bibinfo {year} {2019})}\BibitemShut {NoStop}%
\bibitem [{\citenamefont {Bello}\ \emph {et~al.}(2020)\citenamefont {Bello}, \citenamefont {Kongsuwan}, \citenamefont {Donegan},\ and\ \citenamefont {Hess}}]{qed2}%
  \BibitemOpen
  \bibfield  {author} {\bibinfo {author} {\bibfnamefont {F.}~\bibnamefont {Bello}}, \bibinfo {author} {\bibfnamefont {N.}~\bibnamefont {Kongsuwan}}, \bibinfo {author} {\bibfnamefont {J.~F.}\ \bibnamefont {Donegan}},\ and\ \bibinfo {author} {\bibfnamefont {O.}~\bibnamefont {Hess}},\ }\href {https://doi.org/https://doi.org/10.1021/acs.nanolett.0c01705} {\bibfield  {journal} {\bibinfo  {journal} {Nano Letters}\ }\textbf {\bibinfo {volume} {20}},\ \bibinfo {pages} {5830} (\bibinfo {year} {2020})}\BibitemShut {NoStop}%
\bibitem [{\citenamefont {Kervinen}\ \emph {et~al.}(2019)\citenamefont {Kervinen}, \citenamefont {Ram{\'\i}rez-Mu{\~n}oz}, \citenamefont {V{\"a}limaa},\ and\ \citenamefont {Sillanp{\"a}{\"a}}}]{qed3}%
  \BibitemOpen
  \bibfield  {author} {\bibinfo {author} {\bibfnamefont {M.}~\bibnamefont {Kervinen}}, \bibinfo {author} {\bibfnamefont {J.~E.}\ \bibnamefont {Ram{\'\i}rez-Mu{\~n}oz}}, \bibinfo {author} {\bibfnamefont {A.}~\bibnamefont {V{\"a}limaa}},\ and\ \bibinfo {author} {\bibfnamefont {M.~A.}\ \bibnamefont {Sillanp{\"a}{\"a}}},\ }\href {https://doi.org/https://doi.org/10.1103/PhysRevLett.123.240401} {\bibfield  {journal} {\bibinfo  {journal} {Physical Review Letters}\ }\textbf {\bibinfo {volume} {123}},\ \bibinfo {pages} {240401} (\bibinfo {year} {2019})}\BibitemShut {NoStop}%
\bibitem [{\citenamefont {Sun}\ \emph {et~al.}(2012)\citenamefont {Sun}, \citenamefont {Ma}, \citenamefont {Wang},\ and\ \citenamefont {Nori}}]{qed4}%
  \BibitemOpen
  \bibfield  {author} {\bibinfo {author} {\bibfnamefont {Z.}~\bibnamefont {Sun}}, \bibinfo {author} {\bibfnamefont {J.}~\bibnamefont {Ma}}, \bibinfo {author} {\bibfnamefont {X.}~\bibnamefont {Wang}},\ and\ \bibinfo {author} {\bibfnamefont {F.}~\bibnamefont {Nori}},\ }\href {https://doi.org/https://doi.org/10.1103/PhysRevA.86.012107} {\bibfield  {journal} {\bibinfo  {journal} {Physical Review A}\ }\textbf {\bibinfo {volume} {86}},\ \bibinfo {pages} {012107} (\bibinfo {year} {2012})}\BibitemShut {NoStop}%
\bibitem [{\citenamefont {Wen}\ \emph {et~al.}(2020)\citenamefont {Wen}, \citenamefont {Ivakhnenko}, \citenamefont {Nakonechnyi}, \citenamefont {Suri}, \citenamefont {Lin}, \citenamefont {Lin}, \citenamefont {Chen}, \citenamefont {Shevchenko}, \citenamefont {Nori},\ and\ \citenamefont {Hoi}}]{qed5}%
  \BibitemOpen
  \bibfield  {author} {\bibinfo {author} {\bibfnamefont {P.}~\bibnamefont {Wen}}, \bibinfo {author} {\bibfnamefont {O.}~\bibnamefont {Ivakhnenko}}, \bibinfo {author} {\bibfnamefont {M.}~\bibnamefont {Nakonechnyi}}, \bibinfo {author} {\bibfnamefont {B.}~\bibnamefont {Suri}}, \bibinfo {author} {\bibfnamefont {J.-J.}\ \bibnamefont {Lin}}, \bibinfo {author} {\bibfnamefont {W.-J.}\ \bibnamefont {Lin}}, \bibinfo {author} {\bibfnamefont {J.}~\bibnamefont {Chen}}, \bibinfo {author} {\bibfnamefont {S.}~\bibnamefont {Shevchenko}}, \bibinfo {author} {\bibfnamefont {F.}~\bibnamefont {Nori}},\ and\ \bibinfo {author} {\bibfnamefont {I.-C.}\ \bibnamefont {Hoi}},\ }\href {https://doi.org/https://doi.org/10.1103/PhysRevB.102.075448} {\bibfield  {journal} {\bibinfo  {journal} {Physical Review B}\ }\textbf {\bibinfo {volume} {102}},\ \bibinfo {pages} {075448} (\bibinfo {year} {2020})}\BibitemShut {NoStop}%
\bibitem [{\citenamefont {Malla}\ and\ \citenamefont {Raikh}(2018)}]{qed6}%
  \BibitemOpen
  \bibfield  {author} {\bibinfo {author} {\bibfnamefont {R.~K.}\ \bibnamefont {Malla}}\ and\ \bibinfo {author} {\bibfnamefont {M.}~\bibnamefont {Raikh}},\ }\href {https://doi.org/https://doi.org/10.1103/PhysRevB.97.035428} {\bibfield  {journal} {\bibinfo  {journal} {Physical Review B}\ }\textbf {\bibinfo {volume} {97}},\ \bibinfo {pages} {035428} (\bibinfo {year} {2018})}\BibitemShut {NoStop}%
\bibitem [{\citenamefont {Malla}\ and\ \citenamefont {Raikh}(2022)}]{Mallacont}%
  \BibitemOpen
  \bibfield  {author} {\bibinfo {author} {\bibfnamefont {R.~K.}\ \bibnamefont {Malla}}\ and\ \bibinfo {author} {\bibfnamefont {M.}~\bibnamefont {Raikh}},\ }\href {https://doi.org/https://doi.org/10.1016/j.physleta.2022.128249} {\bibfield  {journal} {\bibinfo  {journal} {Physics Letters A}\ }\textbf {\bibinfo {volume} {445}},\ \bibinfo {pages} {128249} (\bibinfo {year} {2022})}\BibitemShut {NoStop}%
\bibitem [{\citenamefont {Ivakhnenko}\ \emph {et~al.}(2023)\citenamefont {Ivakhnenko}, \citenamefont {Shevchenko},\ and\ \citenamefont {Nori}}]{Oleh1}%
  \BibitemOpen
  \bibfield  {author} {\bibinfo {author} {\bibfnamefont {O.~V.}\ \bibnamefont {Ivakhnenko}}, \bibinfo {author} {\bibfnamefont {S.~N.}\ \bibnamefont {Shevchenko}},\ and\ \bibinfo {author} {\bibfnamefont {F.}~\bibnamefont {Nori}},\ }\href {https://doi.org/https://doi.org/10.1016/j.physrep.2022.10.002} {\bibfield  {journal} {\bibinfo  {journal} {Physics Reports}\ }\textbf {\bibinfo {volume} {995}},\ \bibinfo {pages} {1} (\bibinfo {year} {2023})},\ \bibinfo {note} {nonadiabatic Landau-Zener-St{\"u}ckelberg-Majorana transitions, dynamics, and interference}\BibitemShut {NoStop}%
\bibitem [{\citenamefont {Longstaff}\ and\ \citenamefont {Graefe}(2019)}]{Longstaff_2019_LZ}%
  \BibitemOpen
  \bibfield  {author} {\bibinfo {author} {\bibfnamefont {B.}~\bibnamefont {Longstaff}}\ and\ \bibinfo {author} {\bibfnamefont {E.-M.}\ \bibnamefont {Graefe}},\ }\href {https://doi.org/10.1103/PhysRevA.100.052119} {\bibfield  {journal} {\bibinfo  {journal} {Physical Review A}\ }\textbf {\bibinfo {volume} {100}},\ \bibinfo {pages} {052119} (\bibinfo {year} {2019})}\BibitemShut {NoStop}%
\bibitem [{\citenamefont {Melanathuru}\ \emph {et~al.}(2022)\citenamefont {Melanathuru}, \citenamefont {Malzard},\ and\ \citenamefont {Graefe}}]{Melanathuru_2022}%
  \BibitemOpen
  \bibfield  {author} {\bibinfo {author} {\bibfnamefont {R.}~\bibnamefont {Melanathuru}}, \bibinfo {author} {\bibfnamefont {S.}~\bibnamefont {Malzard}},\ and\ \bibinfo {author} {\bibfnamefont {E.-M.}\ \bibnamefont {Graefe}},\ }\href {https://doi.org/10.1103/PhysRevA.106.012208} {\bibfield  {journal} {\bibinfo  {journal} {Physical Review A}\ }\textbf {\bibinfo {volume} {106}},\ \bibinfo {pages} {012208} (\bibinfo {year} {2022})}\BibitemShut {NoStop}%
\bibitem [{\citenamefont {Kam}\ and\ \citenamefont {Chen}(2023)}]{Kam_2023_LZ}%
  \BibitemOpen
  \bibfield  {author} {\bibinfo {author} {\bibfnamefont {C.-F.}\ \bibnamefont {Kam}}\ and\ \bibinfo {author} {\bibfnamefont {Y.}~\bibnamefont {Chen}},\ }\href {https://arxiv.org/abs/2301.04816} {\bibfield  {journal} {\bibinfo  {journal} {arXiv:2301.04816}\ } (\bibinfo {year} {2023})}\BibitemShut {NoStop}%
\bibitem [{\citenamefont {Demkov}\ and\ \citenamefont {Ostrovsky}(2000)}]{Dem}%
  \BibitemOpen
  \bibfield  {author} {\bibinfo {author} {\bibfnamefont {Y.~N.}\ \bibnamefont {Demkov}}\ and\ \bibinfo {author} {\bibfnamefont {V.~N.}\ \bibnamefont {Ostrovsky}},\ }\href {https://doi.org/10.1103/PhysRevA.61.032705} {\bibfield  {journal} {\bibinfo  {journal} {Physical Review A}\ }\textbf {\bibinfo {volume} {61}},\ \bibinfo {pages} {032705} (\bibinfo {year} {2000})}\BibitemShut {NoStop}%
\bibitem [{\citenamefont {Witthaut}\ \emph {et~al.}(2006)\citenamefont {Witthaut}, \citenamefont {Graefe},\ and\ \citenamefont {Korsch}}]{ICA}%
  \BibitemOpen
  \bibfield  {author} {\bibinfo {author} {\bibfnamefont {D.}~\bibnamefont {Witthaut}}, \bibinfo {author} {\bibfnamefont {E.~M.}\ \bibnamefont {Graefe}},\ and\ \bibinfo {author} {\bibfnamefont {H.~J.}\ \bibnamefont {Korsch}},\ }\href {https://doi.org/10.1103/PhysRevA.73.063609} {\bibfield  {journal} {\bibinfo  {journal} {Physical Review A}\ }\textbf {\bibinfo {volume} {73}},\ \bibinfo {pages} {063609} (\bibinfo {year} {2006})}\BibitemShut {NoStop}%
\bibitem [{\citenamefont {Sinitsyn}(2004)}]{nogo-LZ}%
  \BibitemOpen
  \bibfield  {author} {\bibinfo {author} {\bibfnamefont {N.}~\bibnamefont {Sinitsyn}},\ }\href {https://iopscience.iop.org/article/10.1088/0305-4470/37/44/016} {\bibfield  {journal} {\bibinfo  {journal} {Journal of Physics A: Mathematical and General}\ }\textbf {\bibinfo {volume} {37}},\ \bibinfo {pages} {10691} (\bibinfo {year} {2004})}\BibitemShut {NoStop}%
\bibitem [{\citenamefont {Sun}\ \emph {et~al.}(2019)\citenamefont {Sun}, \citenamefont {Chernyak}, \citenamefont {Piryatinski},\ and\ \citenamefont {Sinitsyn}}]{DPA}%
  \BibitemOpen
  \bibfield  {author} {\bibinfo {author} {\bibfnamefont {C.}~\bibnamefont {Sun}}, \bibinfo {author} {\bibfnamefont {V.~Y.}\ \bibnamefont {Chernyak}}, \bibinfo {author} {\bibfnamefont {A.}~\bibnamefont {Piryatinski}},\ and\ \bibinfo {author} {\bibfnamefont {N.~A.}\ \bibnamefont {Sinitsyn}},\ }\href {https://doi.org/10.1103/PhysRevLett.123.123605} {\bibfield  {journal} {\bibinfo  {journal} {Physical Review Letters}\ }\textbf {\bibinfo {volume} {123}},\ \bibinfo {pages} {123605} (\bibinfo {year} {2019})}\BibitemShut {NoStop}%
\bibitem [{\citenamefont {Sinitsyn}(2015{\natexlab{a}})}]{sinitsyn_exact_2015}%
  \BibitemOpen
  \bibfield  {author} {\bibinfo {author} {\bibfnamefont {N.~A.}\ \bibnamefont {Sinitsyn}},\ }\href {https://doi.org/10.1088/1751-8113/48/19/195305} {\bibfield  {journal} {\bibinfo  {journal} {Journal of Physics A: Mathematical and Theoretical}\ }\textbf {\bibinfo {volume} {48}},\ \bibinfo {pages} {195305} (\bibinfo {year} {2015}{\natexlab{a}})}\BibitemShut {NoStop}%
\bibitem [{\citenamefont {Sinitsyn}(2015{\natexlab{b}})}]{Sinitsyn_2015_four}%
  \BibitemOpen
  \bibfield  {author} {\bibinfo {author} {\bibfnamefont {N.~A.}\ \bibnamefont {Sinitsyn}},\ }\href {https://doi.org/10.1103/PhysRevB.92.205431} {\bibfield  {journal} {\bibinfo  {journal} {Physical Review B}\ }\textbf {\bibinfo {volume} {92}},\ \bibinfo {pages} {205431} (\bibinfo {year} {2015}{\natexlab{b}})}\BibitemShut {NoStop}%
\bibitem [{\citenamefont {Ginzel}\ \emph {et~al.}(2020)\citenamefont {Ginzel}, \citenamefont {Mills}, \citenamefont {Petta},\ and\ \citenamefont {Burkard}}]{petta1}%
  \BibitemOpen
  \bibfield  {author} {\bibinfo {author} {\bibfnamefont {F.}~\bibnamefont {Ginzel}}, \bibinfo {author} {\bibfnamefont {A.~R.}\ \bibnamefont {Mills}}, \bibinfo {author} {\bibfnamefont {J.~R.}\ \bibnamefont {Petta}},\ and\ \bibinfo {author} {\bibfnamefont {G.}~\bibnamefont {Burkard}},\ }\href {https://doi.org/10.1103/PhysRevB.102.195418} {\bibfield  {journal} {\bibinfo  {journal} {Physical Review B}\ }\textbf {\bibinfo {volume} {102}},\ \bibinfo {pages} {195418} (\bibinfo {year} {2020})}\BibitemShut {NoStop}%
\bibitem [{\citenamefont {Mi}\ \emph {et~al.}(2018)\citenamefont {Mi}, \citenamefont {Kohler},\ and\ \citenamefont {Petta}}]{petta2}%
  \BibitemOpen
  \bibfield  {author} {\bibinfo {author} {\bibfnamefont {X.}~\bibnamefont {Mi}}, \bibinfo {author} {\bibfnamefont {S.}~\bibnamefont {Kohler}},\ and\ \bibinfo {author} {\bibfnamefont {J.~R.}\ \bibnamefont {Petta}},\ }\href {https://doi.org/10.1103/PhysRevB.98.161404} {\bibfield  {journal} {\bibinfo  {journal} {Physical Review B}\ }\textbf {\bibinfo {volume} {98}},\ \bibinfo {pages} {161404} (\bibinfo {year} {2018})}\BibitemShut {NoStop}%
\bibitem [{\citenamefont {Malla}\ and\ \citenamefont {Raikh}(2017)}]{Malla2017nonint}%
  \BibitemOpen
  \bibfield  {author} {\bibinfo {author} {\bibfnamefont {R.~K.}\ \bibnamefont {Malla}}\ and\ \bibinfo {author} {\bibfnamefont {M.~E.}\ \bibnamefont {Raikh}},\ }\href {https://doi.org/10.1103/PhysRevB.96.115437} {\bibfield  {journal} {\bibinfo  {journal} {Physical Review B}\ }\textbf {\bibinfo {volume} {96}},\ \bibinfo {pages} {115437} (\bibinfo {year} {2017})}\BibitemShut {NoStop}%
\bibitem [{\citenamefont {Volkov}\ and\ \citenamefont {Ostrovsky}(2005)}]{Volkov_2005}%
  \BibitemOpen
  \bibfield  {author} {\bibinfo {author} {\bibfnamefont {M.~V.}\ \bibnamefont {Volkov}}\ and\ \bibinfo {author} {\bibfnamefont {V.~N.}\ \bibnamefont {Ostrovsky}},\ }\href {https://doi.org/10.1088/0953-4075/38/7/011} {\bibfield  {journal} {\bibinfo  {journal} {Journal of Physics B: Atomic, Molecular and Optical Physics}\ }\textbf {\bibinfo {volume} {38}},\ \bibinfo {pages} {907} (\bibinfo {year} {2005})}\BibitemShut {NoStop}%
\bibitem [{\citenamefont {Malla}\ \emph {et~al.}(2021)\citenamefont {Malla}, \citenamefont {Chernyak},\ and\ \citenamefont {Sinitsyn}}]{malla2021}%
  \BibitemOpen
  \bibfield  {author} {\bibinfo {author} {\bibfnamefont {R.~K.}\ \bibnamefont {Malla}}, \bibinfo {author} {\bibfnamefont {V.~Y.}\ \bibnamefont {Chernyak}},\ and\ \bibinfo {author} {\bibfnamefont {N.~A.}\ \bibnamefont {Sinitsyn}},\ }\href {https://doi.org/10.1103/PhysRevB.103.144301} {\bibfield  {journal} {\bibinfo  {journal} {Physical Review B}\ }\textbf {\bibinfo {volume} {103}},\ \bibinfo {pages} {144301} (\bibinfo {year} {2021})}\BibitemShut {NoStop}%
\bibitem [{\citenamefont {Chernyak}\ and\ \citenamefont {Sinitsyn}(2021)}]{Chernyak_2021}%
  \BibitemOpen
  \bibfield  {author} {\bibinfo {author} {\bibfnamefont {V.~Y.}\ \bibnamefont {Chernyak}}\ and\ \bibinfo {author} {\bibfnamefont {N.~A.}\ \bibnamefont {Sinitsyn}},\ }\href {https://doi.org/10.1088/1751-8121/abe427} {\bibfield  {journal} {\bibinfo  {journal} {Journal of Physics A: Mathematical and Theoretical}\ }\textbf {\bibinfo {volume} {54}},\ \bibinfo {pages} {115204} (\bibinfo {year} {2021})}\BibitemShut {NoStop}%
\bibitem [{\citenamefont {Dykhne}(1962)}]{dykhne}%
  \BibitemOpen
  \bibfield  {author} {\bibinfo {author} {\bibfnamefont {A.~M.}\ \bibnamefont {Dykhne}},\ }\href {http://jetp.ras.ru/cgi-bin/dn/e_014_04_0941.pdf} {\bibfield  {journal} {\bibinfo  {journal} {Soviet Physics - Journal of Experimental and Theoretical Physics}\ }\textbf {\bibinfo {volume} {14}},\ \bibinfo {pages} {1} (\bibinfo {year} {1962})}\BibitemShut {NoStop}%
\bibitem [{\citenamefont {Erdelyi}(1953)}]{parabolic}%
  \BibitemOpen
  \bibinfo {editor} {\bibfnamefont {A.}~\bibnamefont {Erdelyi}},\ ed.,\ \href@noop {} {\emph {\bibinfo {title} {Higher Transcendental Functions Volume 2}}}\ (\bibinfo  {publisher} {McGraw-Hill, New York},\ \bibinfo {year} {1953})\BibitemShut {NoStop}%
\bibitem [{\citenamefont {Malla}\ \emph {et~al.}(2017)\citenamefont {Malla}, \citenamefont {Mishchenko},\ and\ \citenamefont {Raikh}}]{malla2017}%
  \BibitemOpen
  \bibfield  {author} {\bibinfo {author} {\bibfnamefont {R.~K.}\ \bibnamefont {Malla}}, \bibinfo {author} {\bibfnamefont {E.~G.}\ \bibnamefont {Mishchenko}},\ and\ \bibinfo {author} {\bibfnamefont {M.~E.}\ \bibnamefont {Raikh}},\ }\href {https://doi.org/10.1103/PhysRevB.96.075419} {\bibfield  {journal} {\bibinfo  {journal} {Physical Review B}\ }\textbf {\bibinfo {volume} {96}},\ \bibinfo {pages} {075419} (\bibinfo {year} {2017})}\BibitemShut {NoStop}%
\end{thebibliography}%

\end{document}